\documentclass[10pt,journal]{IEEEtran}

\usepackage{graphics}
\usepackage{graphicx}
\usepackage{caption}
\usepackage{subcaption}
\usepackage[T1]{fontenc}
\usepackage[numbers,sort&compress]{natbib}
\usepackage[ruled,vlined]{algorithm2e}
\usepackage{float}
\usepackage{amsfonts}
\usepackage{amsmath}
\usepackage{amssymb}
\usepackage{url}
\usepackage{enumitem}
\usepackage{multirow}
\usepackage[table,xcdraw]{xcolor}
\usepackage{footnote}
\usepackage[symbol]{footmisc}
\usepackage{tikz,xcolor}
\usepackage[hidelinks]{hyperref}
\usepackage{cleveref}

\SetAlgorithmName{Procedure}{procedure}{List of Procedures}

\definecolor{lime}{HTML}{A6CE39}
\DeclareRobustCommand{\orcidicon}{%
	\begin{tikzpicture}
	\draw[lime, fill=lime] (0,0) 
	circle [radius=0.16] 
	node[white] {{\fontfamily{qag}\selectfont \tiny ID}};
	\draw[white, fill=white] (-0.0625,0.095) 
	circle [radius=0.007];
	\end{tikzpicture}
	\hspace{-2mm}
}

\foreach \x in {A, ..., Z}{%
	\expandafter\xdef\csname orcid\x\endcsname{\noexpand\href{https://orcid.org/\csname orcidauthor\x\endcsname}{\noexpand\orcidicon}}
}

\begin{document}

\title{\textit{FairShare}: Blockchain Enabled Fair, Accountable and Secure Data Sharing for Industrial IoT}

\author{Jayasree~Sengupta\orcidA{},
        Sushmita~Ruj\orcidB{},~\IEEEmembership{Senior~Member,~IEEE,}
        and~Sipra~Das~Bit\orcidC{},~\IEEEmembership{Senior~Member,~IEEE}
\thanks{J. Sengupta and S. Das Bit are with the Department of Computer Science and Technology, Indian Institute of Engineering Science and Technology, Howrah, India, E-Mail: (jayasree202@gmail.com ; sdasbit@yahoo.co.in).}
\thanks{S. Ruj is with School of Computer Science and Engineering, University of New South Wales, Sydney, Australia, E-Mail: (sushmita.ruj@unsw.edu.au).}
}

\maketitle

\begin{abstract}
Industrial Internet of Things (IIoT) opens up a challenging research area towards improving secure data sharing which currently has several limitations. Primarily, the lack of inbuilt guarantees of honest behavior of participating, such as end-users or cloud behaving maliciously may result in disputes. Given such challenges, we propose a fair, accountable, and secure data sharing scheme, \textit{FairShare} for IIoT. In this scheme, data collected from IoT devices are processed and stored in cloud servers with intermediate fog nodes facilitating computation. Authorized clients can access this data against some fee to make strategic decisions for improving the operational services of the IIoT system. By enabling blockchain, \textit{FairShare} prevents fraudulent activities and thereby achieves fairness such that each party gets their rightful outcome in terms of data or penalty/rewards while simultaneously ensuring accountability of the services provided by the parties. Additionally, smart contracts are designed to act as a mediator during any dispute by enforcing payment settlement. Further, security and privacy of data are ensured by suitably applying cryptographic techniques like proxy re-encryption. We prove \textit{FairShare} to be secure as long as at least one of the parties is honest. We validate \textit{FairShare} with a theoretical overhead analysis. We also build a prototype in Ethereum to estimate performance and justify comparable results with a state-of-the-art scheme both via simulation and a realistic testbed setup. We observe an additional communication overhead of 256 bytes and a cost of deployment of 1.01 USD in Ethereum which are constant irrespective of file size.
\end{abstract}


\section{Introduction}

Industry 4.0 focuses on the automation of manufacturing industries \cite{i3}, whereas Industrial Internet of Things (IIoT) refers to its non-manufacturing counterparts. The ever-increasing heterogeneous industrial data, the unpredictable network latency and the high cost of communication bandwidth, renders it almost impractical for traditional cloud-based systems to meet the stringent latency requirements of controlling critical industrial applications. Thus, IIoT architecture is evolving from remote cloud to proximity fog to facilitate automation, monitoring and diagnosis in smart manufacturing \cite{i2,i4}. 

The wide range of benefits (e.g. scalability, automation, decentralisation) offered by these IIoT systems comes at the cost of high data generation  from various physical components such as human-machine interface, assembly lines etc \cite{i6}. However, these data collected maybe useful for different stakeholders to increase efficiency and maintain smooth productivity within the industry. For instance, production rate of particular machinery can be of interest to multiple end-users including both manufacturers and suppliers. Thus, for an industrial setting, vast amount of sensitive data need to be outsourced to the cloud for sharing among various users for analysis purposes. 

The traditional data sharing methods either have huge computation/bandwidth overhead \cite{XWZ_ASIACCS'12} or lack efficient user management schemes \cite{r12,r21,r13,r18,r22,r14}. By using such methods, the sender either needs to encrypt the data individually for each user or update encryption keys and re-distribute them every time a user joins/leaves the system. Present day systems \cite{r5,r6,r7,r9,r10,r23,r25,r26} have been able to solve a few of these issues, but they still lack the ability to guarantee honest behaviour of the participating parties forcing them to trust the data they receive. For example, end-users trust the cloud to deliver the right data/files against payment. Similarly, the cloud expects the end-users to pay for a successful delivery, without falsely alleging misinformation or non-delivery. Thus, the lack of inbuilt guarantees of honest behaviour can lead to disputes such as, the end-user stealing the content by maliciously claiming the non-receipt of data, thereby denying payment; or cloud charging the end-users without providing them the correct data. Lastly, the involved parties: Fog, Cloud and Client being strategic players would often collude and cheat to gain more money. Hence, in order to design a fair, accountable and secure data sharing scheme, the following are the concerns:

\begin{itemize}[leftmargin=*]
    \item Establish an agreement between the participating parties such as Fog, Cloud and the end-users (i.e. Client) to ensure:
    \begin{itemize}[leftmargin=*]
        \item security, privacy, accountability and fairness of the data exchanged and the services provided.
        \item honest behaviour by penalizing the fraudulent activities which guarantees trustworthiness and transparency.
        \item each party gets their rightful outcome (i.e. data and/or penalty/rewards) or no one does.
    \end{itemize}
    \item Detect a malicious party as well as colluding parties in case of any corrupt behaviour.
    \item Provide performance efficiency in terms of:
    \begin{itemize}[leftmargin=*]
        \item reasonable computation overhead while still maintaining the security aspects.
        \item facilitating new user addition/revocation.
    \end{itemize}
\end{itemize}

Proxy re-encryption is an attractive option for privacy preserving data sharing  because it requires the sender to generate only single/multiple re-encryption key/s for new user addition or for sharing data with multiple recipients. We know that fair exchange between two parties is not possible without a trusted set up \cite{PG99}. By fairness, we mean that each party gets their rightful outcome (i.e. data/penalty/rewards). Here, blockchain acts as this trusted intermediary and resolves dispute that might arise because of non-payment of fees or incorrect data. Blockchain also helps to settle payments and keeps a record of transactions to maintain transparency.

\subsection{Major Contributions}

\noindent The contributions put forth by our article are as follows:

\begin{enumerate}[leftmargin=*]
    \item Propose a fair, accountable and secure data sharing scheme named $\textit{FairShare}$ where:
    \begin{enumerate}[leftmargin=*]
        \item proxy re-encryption has been suitably applied to ensure security and privacy of data exchanged.
        \item smart contracts are designed and carefully integrated into the blockchain which guarantees accountability and fairness by preventing fraudulent activities and resolving disputes amongst parties.
       
    \end{enumerate}
    \item Establish provable security guarantees even when participating parties collude with one another.
    \item Validate $\textit{FairShare}$ with a theoretical overhead analysis and practical prototype implementation in Ethereum including simulation and testbed experimentation by:
    \begin{enumerate}[leftmargin=*]
        \item evaluating computation time, communication overhead and cost of deployment which are rather small for file sizes as large as 10MB.
        \item justifying comparable performance with a competitor via simulation where data storage time is a few seconds and data retrieval is close to a minute.
        \item establishing scalability with increasing number of devices in the network through testbed experimentation.
    \end{enumerate}
\end{enumerate}

\subsection{Organization}

The rest of the paper is structured as follows. Section \ref{RW} gives an overview of the current state-of-the art works. Section \ref{Section2} briefly describes the basic building blocks. Section \ref{Section3} gives an overview of the system model. Our proposed scheme is described in Section \ref{Sec4}. The security analysis of our proposed scheme is presented in Section \ref{Section5}. Section \ref{Section6} discusses the implementation along with the performance analysis results. Section \ref{Section7} finally concludes our work.

\section{Related Work \label{RW}}

This section provides a brief overview on the various secure data sharing schemes relevant to IoT and IIoT that fosters sharing data securely amongst various parties. Next, we explain dispute resolution protocols that serve as a mediator to resolve disputes amongst participating parties. Lastly, some blockchain specific re-encryption techniques are presented that suitably utilise blockchain to enable secure data sharing.

\noindent\textbf{Secure Data Sharing in IoT:} The works \cite{r15,r16,r17} have proposed various frameworks, middleware platforms for building an efficient and decentralised IoT data marketplace. Their primary focus is to exchange real-time IoT data between entities by leveraging smart contract based agreements. Contrarily, Xu et al. \cite{r24} have proposed a blockchain-based secure, flexible data sharing platform with fine-grained access control.\\
\noindent\textbf{Secure Data Sharing in IIoT:} In \cite{r12,r21} the authors focus on developing blockchain-based secure data exchange mechanisms among smart entities in IIoT. Contrarily, the work \cite{r13} has proposed a differential privacy based data sharing framework in the presence of multiple consumers. For energy trading in smart grids and secure data sharing, the works \cite{r18,r22} have proposed a Secure Private Blockchain (SPB) based framework and a Directed Acyclic Graph based V2G network using blockchain respectively. Lastly, the work \cite{r14} has proposed a blockchain empowered privacy preserving machine learning based approach for efficient data sharing among multiple distributed parties. However, all of these schemes face a few major drawbacks like difficulty in modification of access policy due to high computation and communication overhead for both data owner and client, data ownership issues, etc. Moreover, they also lack the ability to capture malicious behaviours of participating parties due to the absence of an indisputable log. These schemes are either dependent on a centralised cloud or they lack the ability to securely share data while guaranteeing fair exchange in case of a dispute.\\
\noindent\textbf{Dispute Resolution Protocols:} To handle dispute resolution, the works \cite{r20,r19,c3,AK_ICBC'19} have proposed two party Ethereum-based frameworks to mediate interactions between parties by consulting a protocol-specific verifier in case of a misbehaviour. These works \cite{r20,r19,c3,AK_ICBC'19} have fairness definitions specific to their application, which is different from our perspective. Specifically, the work \cite{c3} stands out amongst the rest, primarily because it has been able to greatly minimize the cost of running smart contracts by avoiding expensive cryptographic tools. Additionally, Singh et al. \cite{SMHK_JPDC'21} have proposed a cross-domain platform to detect malicious parties and take necessary actions in case of dispute. However, all of these works are based on exchanging data between two parties which is entirely different from our objective of sharing data with multiple end-users while simultaneously reducing the overheads.\\
\noindent\textbf{Blockchain-based Proxy Re-Encryption:} To tackle the challenges faced by the traditional data sharing schemes, recently researchers have discussed the need to implement blockchain as an underlying infrastructure in various applications like smart grid, healthcare etc. \cite{r27}. Therefore, the state-of-the art data sharing schemes \cite{r5,r6,r7,r9,r10,r23,r25,r26} have introduced blockchain as a backbone with proxy re-encryption for achieving data confidentiality, decentralized data storage, flexible user management. Nevertheless, they are either based on the assumption that different parties are not colluding with one another or they do not consider agreements between parties for rightfully paying/penalizing them in order to ensure fairness. 

\begin{table}[!t]
\centering
\caption{\small \sl Overview on Current Literature}
\label{tab:Table10}
\scalebox{0.63}{%
\begin{tabular}{|l|l|l|l|}
\hline
\multicolumn{1}{|c|}{\textbf{Area of Research}} & \multicolumn{1}{c|}{\textbf{Citation}} & \multicolumn{1}{c|}{\textbf{Contribution}} & \multicolumn{1}{c|}{\textbf{Research Gap}} \\ \hline
\begin{tabular}[c]{@{}l@{}}Secure Data\\ Sharing in IoT\end{tabular} & \cite{r15,r16,r17,r24} & \begin{tabular}[c]{@{}l@{}}Framework or middleware\\ platform for decentralised\\ IoT data marketplace\end{tabular} & \begin{tabular}[c]{@{}l@{}} Industrial application (e.g. smart grid, smart\\ manufacturing) specific vulnerabilities like \\ secure energy trading, secure data sharing \\ with users respectively remain unaddressed \end{tabular}  \\ \hline
\begin{tabular}[c]{@{}l@{}}Secure Data\\ Sharing in IIoT\end{tabular} & \cite{r12,r21,r13,r18,r22,r14} & \begin{tabular}[c]{@{}l@{}}Blockchain-based data sharing;\\ Differential privacy-based data\\ sharing; Secure Private Blockchain\\ (SPB); Blochain-based ML\end{tabular} & \begin{tabular}[c]{@{}l@{}}Difficulty in access policy modification;\\High computation and  communication\\overheads; Lack the ability to capture\\ malicious behaviour of participating parties \end{tabular} \\ \hline
\begin{tabular}[c]{@{}l@{}}Dispute Resolution\\ Protocols\end{tabular} & \cite{r20,r19,c3,SMHK_JPDC'21} & \begin{tabular}[c]{@{}l@{}}Two party Ethereum-based\\ framework; Cross domain platform\\ using blockchain\end{tabular} & \begin{tabular}[c]{@{}l@{}}Doesn't address dispute resolution during\\data sharing with multiple end-users\end{tabular} \\ \hline
\begin{tabular}[c]{@{}l@{}}Blockchain-based\\ Proxy Re-Encryption\end{tabular} & \cite{r5,r6,r7,r9,r10,r23,r25,r26} & \begin{tabular}[c]{@{}l@{}}Achieves data confidentiality,\\ decentralised data storage, flexible\\ user management\end{tabular} & \begin{tabular}[c]{@{}l@{}}Cannot handle collusion cases between\\ parties; Doesn't ensure fairness for\\rightfully paying/penalizing parties\end{tabular} \\ \hline
\end{tabular}%
}
\end{table}

Table \ref{tab:Table10} summarizes the contributions of the current literature. From the table, we observe the potential research gap which motivates us to propose a blockchain-based fair, accountable and secure data sharing scheme to eliminate the existing drawbacks while realising that same data maybe shared with multiple end-users. Also, developing our scheme on a Fog-based platform, helps us utilize the benefits of the architecture. In this way, we fulfil the requirements outlined earlier by delegating key management and re-encryption tasks to fog nodes \cite{i2} thereby reducing the dependence on cloud.

\section{Preliminaries \label{Section2}}

The cryptographic building blocks used in our scheme are:

\subsection{Bilinear Maps and Pairing Concept}

Two groups $\mathbb{G}_1$ and $\mathbb{G}_T$ of large prime order \textit{q} is chosen where $g$ is the generator of $\mathbb{G}_1$. The bilinear pairing is defined as a map $\hat{e} : \mathbb{G}_1 \times \mathbb{G}_1 \rightarrow \mathbb{G}_T$ \cite{c10}. Here, $\mathbb{G}_1$ is a subgroup of the additive group of points on an elliptic curve $E/\mathbb{F}_p$. $\mathbb{G}_T$ is a
subgroup of the multiplicative group of a finite field $\mathbb{F}^*_{p^2}$ \cite{c4}. The map $\hat{e}$ has the following properties:

\begin{itemize}[leftmargin=*]
\item \textbf{Bilinear:} For all P, Q $\in$ $\mathbb{G}_1$ and $\forall$ c,d $\in$ $\mathbb{Z}^*_q$, we have $\hat{e}(cP,dQ) = \hat{e}(cP,Q)^d = \hat{e}(P,dQ)^c = \hat{e}(P,Q)^{cd}$.

\item \textbf{Non-degenerate:} If P is a generator of $\mathbb{G}_1$, then
$\forall P \in \mathbb{G}_1, P \neq 0 \Rightarrow \ \hat{e}(P,P) = \mathbb{G}_T$ [i.e. $\hat{e}\ (P,P)$ generates $\mathbb{G}_T$].

\item \textbf{Computable:} There is an efficient algorithm to compute $\hat{e}(P,Q)$  $\forall \ P,Q \in \mathbb{G}_1$.

\end{itemize}

In our work, bilinear pairing acts as the fundamental concept behind the proxy re-encryption scheme described below.

\subsection{Proxy Re-Encryption \label{Sec2.4}}

Re-Encryption is a mechanism which allows a proxy (typically a third party) to alter (i.e. re-encrypt) an encrypted message from a sender in a way that it can be decrypted by the receiver using his/her private key \cite{c2}. A unidirectional, single-hop proxy re-encryption scheme is chosen consisting of a tuple of algorithms defined as below :

\begin{itemize}[leftmargin=*]
    \item $PRE.KeyGen\ (\lambda , par) \rightarrow (sk_{id} , pk_{id}$) : Using public parameter $par$ and security parameter $\lambda$ as input to this randomized function, each of the involved parties derive a private/public key-pair.
    \item $PRE.ReKeyGen\ (par , sk_i , pk_j) \rightarrow (rk_{i \rightarrow j}$) : Generates a re-encryption key ($rk_{i \rightarrow j}$) which can convert a ciphertext intended for user $i$ (Sender) into a ciphertext intended for user $j$ (Receiver).
    \item $PRE.Enc\ (par , pk_i , m) \rightarrow (c'$) : Encrypts a message $m$ to generate ciphertext $c'$, which can be re-encrypted for an intended recipient.
     \item $PRE.ReEnc\ (par , rk_{i \rightarrow j} , c') \rightarrow (c''$) : Re-encrypts $c'$ to generate ciphertext $c''$, which can easily be decrypted by the intended recipient using his/her private key.
     \item $PRE.Dec\ (par , sk_j , c'') \rightarrow (m$) : Outputs $m$ or generates an \textit{\textbf{`invalid'}} message on input of $c''$.
\end{itemize}

In our proposed scheme, re-encryption has been used to securely share the file encryption key with multiple end-users.

\subsection{Blockchain}

A blockchain is a tamper-resistant, immutable, distributed verifiable ledger. Information stored in blockchain is made of a chain of blocks where each block consists of a series of transactions. The blocks are typically hash-linked in such a way, that if a transaction is modified in one block it has to be altered in all the subsequent blocks. A smart contract is a self-enforcing piece of computer program that can be used to formalize simple agreements between two parties and control the transfer of digital currencies or assets between them \cite{c6}. In smart contracts, functions can be defined beyond the exchange of cryptocurrencies, such as access policy verification.

\noindent Ethereum \cite{c8} is an open-source popular public blockchain based distributed computing platform which supports Turing-complete languages to feature smart contract functionality. Smart contracts \cite{c17} in Ethereum are written in a scripting language like Solidity. Once a contract is deployed its execution can be triggered via transactions, which are processed by miners \cite{c3}, who are special nodes responsible for validating and adding transactions to blocks \cite{c5}. Ether, the native cryptocurrency of Ethereum is used to incentivize miners to execute smart contracts by processing transactions. The number of operations executed is used to decide the transaction fees which is then paid in gas, an internal Ethereum currency.

\section{System Model \label{Section3}}

The section illustrates our system model where we adopt the architecture proposed in \cite{i2} as its backbone. 

\subsection{Components}

The model consists of five major entities as enlisted below.

\begin{itemize}[leftmargin=*]
    \item \textbf{Industrial IoT (IIoT) Device (D):} They are any form of industrial equipment like Automated Guided Vehicles (AGVs) embedded with different types of sensors (temperature, pressure etc.) \cite{WSB_ANTS'19} to enable a smart infrastructure. 
    \item \textbf{Fog Node (F):} They act as a potential middleware between resource constrained IIoT devices and cloud servers. Fog nodes process raw data collected from IIoT devices and secure them using the underlying blockchain platform. Smart devices with higher computational power like smartphones/tablets, laptops act as fog nodes in our system \cite{SRB_CCGrid'22}.
    \item \textbf{Blockchain Network (B):} Our proposed model uses a public blockchain like Ethereum \cite{c8} which is basically a permissionless blockchain that allows anyone to join the network. The participants can read and write on to the blockchain where the data can't be changed once it is validated \cite{c16}. The critical operations within the blockchain (fair exchange of data) are governed by smart contracts which are also globally accessible \cite{c5}. 
    \item \textbf{Cloud (Cl):} It is responsible for storing huge amount of raw/processed data collected by IIoT devices and further processed by fog. Cloud also processes data access requests from clients with the help of underlying blockchain network.
    \item \textbf{Client (C):} In an industrial setting, clients can typically be manufactures, suppliers or customers who remotely request/access data via some telecommunication devices.
\end{itemize}

\begin{figure} [!t]
\begin{center}  
\fbox{\includegraphics[scale=0.40]{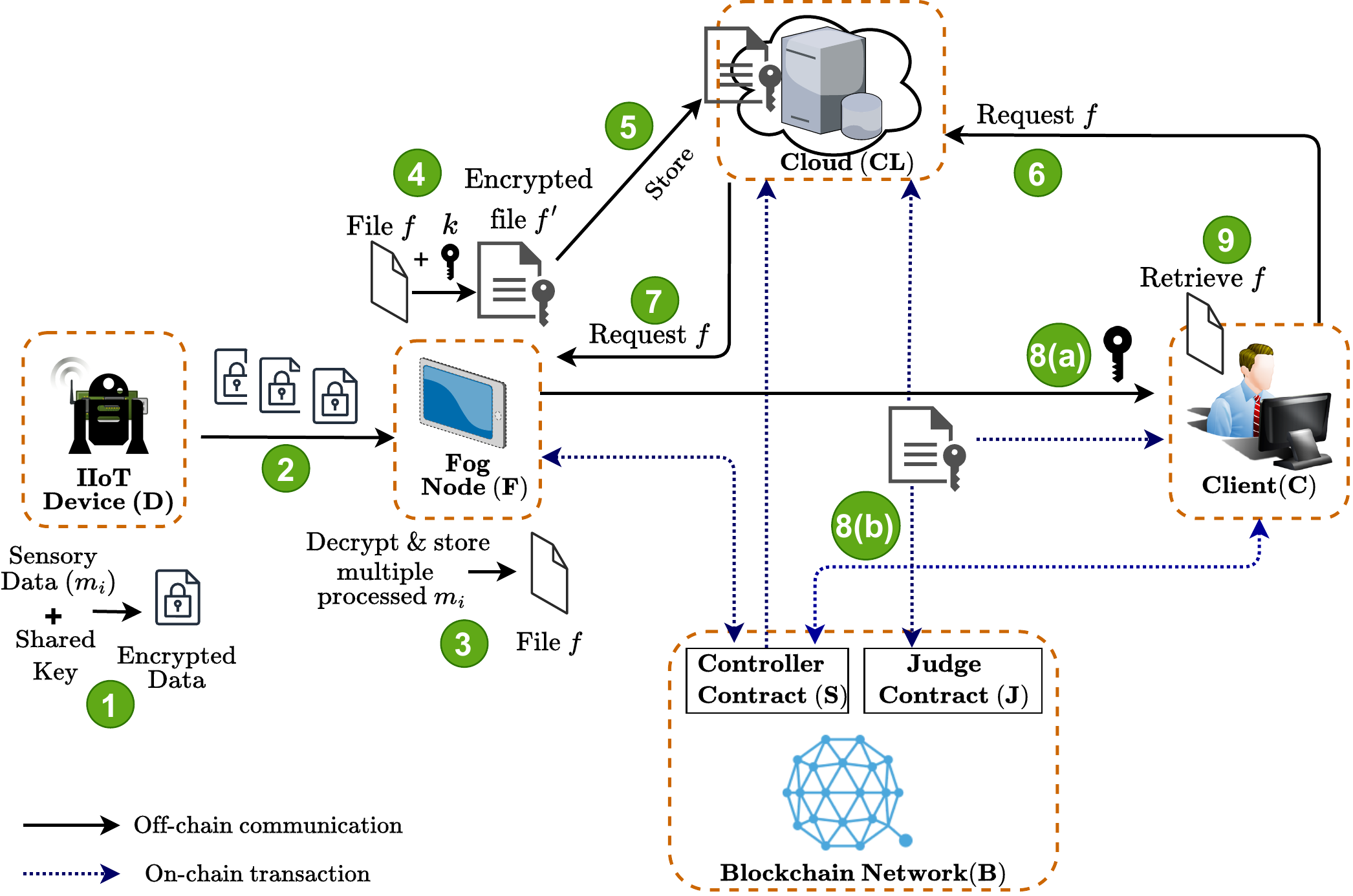}}
\setlength{\belowcaptionskip}{-10pt}
\caption{\small \sl Overview of Interactions amongst Parties} \label{fig:Image1}
\end{center}  
\end{figure}

\noindent Fig. \ref{fig:Image1} shows a brief overview of the interactions between parties. Here, sensed data $m_i$ from an IIoT device is encrypted and sent to fog node F. In this context, a couple of works \cite{HET_BC'19,PV_ICC'20} have proposed solutions to securely migrate data from external sources (e.g. IoT devices) to the blockchain network. Those works either integrate blockchain at the device level or build a trustworthy system to reliably transfer data to the blockchain. However, in \textit{FairShare} to reduce data redundancy and overheads on the blockchain network, data received from IIoT devices are processed by F before migrating to the blockchain. The fog decrypts the data and stores summaries of such data to a file $f$. This file is then encrypted to $f'$ and stored in the cloud for further processing. It is quite logical that in such a scenario, fog being less resource-constrained will perform the necessary interactions with the blockchain. Hence, unlike \cite{HET_BC'19,PV_ICC'20}, there are no direct interactions between the IIoT devices and the blockchain in our proposal as shown in Fig. \ref{fig:Image1}. However, all necessary measures (e.g. preserving authenticity) for securely transferring data from IIoT devices to fog nodes are taken care of by \textit{FairShare}. Finally, in this setup, Client (C) issues request for non real-time data only. The cloud on receiving such a request forwards it to F. The cloud and fog then concurrently interacts with the blockchain platform via smart contracts to securely share encrypted file $f'$ and the necessary key to the client for decryption. C then decrypts $f'$ using this key to retrieve $f$.

\subsection{Security Guarantees and Adversarial Model \label{Sec3.2}}

The security features, adversarial models and assumptions considered in \textit{FairShare} are discussed here.

\subsubsection{Security Requirements}

\noindent The security requirements addressed by the scheme are:

\begin{itemize}[leftmargin=*]
     \item \textbf{Authenticity:} The authenticity of the data exchanged (file $f$) requires that none of the parties (Fog or Cloud) can forge a valid $f$ in a way to fool an honest recipient \cite{c5}. This is only possible with a negligible probability in $\lambda$, if they are able to find a collision in the hash function.
    \item \textbf{Privacy:} The privacy of data stored in cloud (file $f'$) requires it to not learn anything about the data or access policies $\rho$ related to $f'$. Cloud only processes access request from client and calls a smart contract to verify its access rights. Cloud shouldn't be able to derive $f$ from the responses obtained.   
    \item \textbf{Accountability and Fairness:} All parties should be accountable for the services they provide. For example, fog and cloud offer services in exchange of payment from client. The fairness property would thus require the following:
    \begin{itemize}[leftmargin=*]
        \item If Cl sends the correct encrypted file ($f'$), then Cl is paid as per the agreements in \textit{FairShare}. On the contrary, if Cl behaves maliciously it gets penalized.
        \item If C is able to retrieve the shared key required to decrypt $f'$, then F receives adequate incentive as per the agreements, otherwise it gets penalized.
        \item If C receives the designated services as intended from both Cl and F, then it has to pay as per the agreements. However, if C incurs any loses due to a malicious participant, then C will be compensated for the damage caused.
    \end{itemize}
\end{itemize}

\subsubsection{Adversarial Model} It consists of the following parties:\\
\noindent \textbf{User/Party:} Any party or user is said to be honest if s/he follows the protocol. If any of them behaves incoherently or deviates arbitrarily, it is said to be malicious.

\noindent \textbf{Adversary:} An adversary is 
a polynomial-time algorithm that can compromise any user at any point of time, subject to some upper bound \cite{c5}. The adversary is assumed to be dynamic in nature where it can make coordinated attacks by sending/receiving messages on behalf of the malicious users. However, the adversary can neither interfere with the message exchanges between honest users/parties nor can it break cryptographic primitives like signatures or hash functions, except with negligible probability. Moreover, the adversary has bounded computational power and storage capabilities.  Lastly, the adversary is considered to be a rational player and it doesn't behave maliciously or corrupt other parties unless it receives sufficient benefit (e.g. incentives/data).

\noindent \textbf{Assumptions:} Based on the current literature \cite{c5,c14,NC_22,AK_ICBC'19}, the following assumptions are considered:

\begin{itemize}[leftmargin=*]
    \item We assume that all devices in this model are functioning as they should (i.e. node capture attacks are infeasible) \cite{NC_22}.
    \item The underlying blockchain platform is secure, having honest majority amongst the peers \cite{c5}.
    \item We do not consider bribery-related attacks \cite{c14}, hence fog nodes cannot sell information to any third party.
    \item Lastly, the initial hash value of file $f$  are agreed upon by all the communicating parties \cite{AK_ICBC'19}.
\end{itemize}

\section{Overview of our Blockchain-based Data Sharing Scheme \label{Sec4}}


\begin{figure*}[htb]
\begin{center}  
\fbox{\includegraphics[scale=0.65]{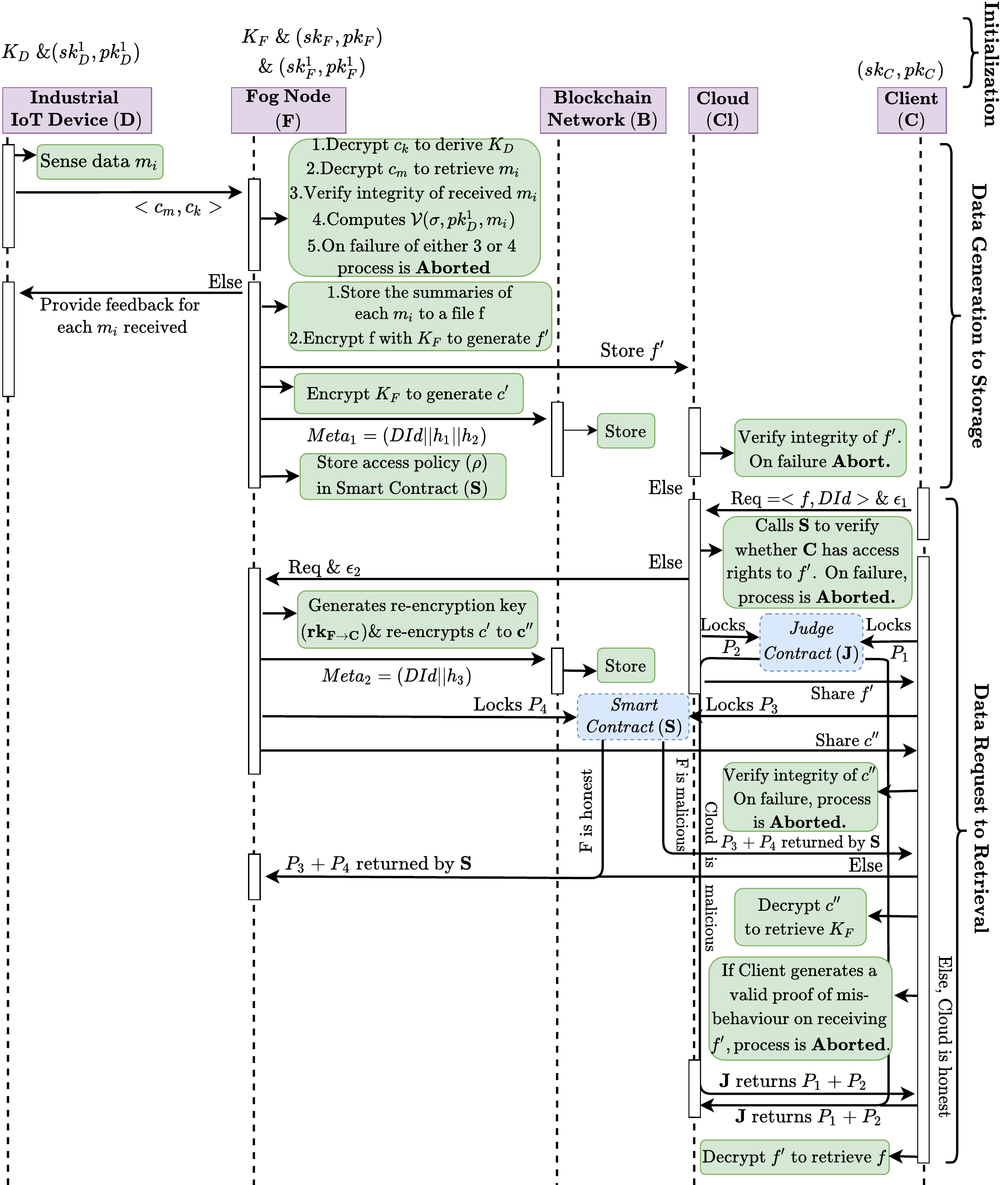}}
\setlength{\belowcaptionskip}{-15pt}
\caption{\small \sl Sequence Diagram of \textit{FairShare} \label{fig:Image2}}
\end{center}  
\end{figure*}

Re-Encryption has been used as the backbone for data sharing, where we suitably modify Proxy Re-encryption scheme \cite{c2} to adopt into our system. This scheme is opted because it is efficient, simple to implement compared to other similar schemes \cite{r1,r2,r4}. The said scheme \cite{c2} uses certain additional parameters in each round of encryption to check validity of the ciphertext. We have suitably modified it \cite{c2} by delegating the validity checking to our inherent blockchain platform. We also integrate other security blocks to develop a full-fledged fair and secure data sharing platform suitable for IIoT scenarios. The misbehaviour detection in \textit{FairShare} is inspired from \textit{FairSwap} \cite{c3} due to its benefits highlighted in Section \ref{RW}.

In \textit{FairShare}, smart contracts are deployed on the blockchain to act as an enforcer of rules. We have designed two contracts: (a) Judge Contract ($\mathbf{J}$) to handle fair exchange of file and (b) Controller Contract ($\mathbf{S}$) to manage other necessary operations. Additionally, in case of any dispute the Turing-completeness of blockchain has been used to codify all necessary actions. Further, the native currency of blockchain is used to control and distribute incentives. The immutability of blockchain has been used to record all data and control message exchanges in the system to provide a transparent infrastructure. Table \ref{tab:Table1} summarizes the notations used.

\begin{table}[!b]
\centering
\caption{\small \sl Notations used in our Scheme}
\label{tab:Table1}
\scalebox{0.80}{
\begin{tabular}{c|c}
\hline
\textbf{Notations} & \textbf{Meaning} \\ \hline
$K_D$/$K_F$ & Symmetric Key generated by D and F respectively \\ \hline
($sk_F , pk_F$) & Key-pair used by F for Re-Encryption \\ \hline
($sk_C , pk_C$) & Key-pair used by C for Re-Encryption \\ \hline
($sk^1_D , pk^1_D$) & Key-pair used by D for Public Key Encryption \\ \hline
($sk^1_F , pk^1_F$) & Key-pair used by F for Public Key Encryption \\ \hline
$H$ & Hash function mapping $\{0,1\}^* \rightarrow \{0,1\}^n$ \\ \hline
$\mathcal{S}$/$\mathcal{V}$ & Signature generation/verification function \\ \hline
$m_i$ & Sensed data \\ \hline
$f$ & File containing summaries of sensed data $m_1,m_2, \ldots  ,m_n$ \\ \hline
\end{tabular}%
}
\end{table}

\noindent \textbf{Working Principle:} Data originating from a specific sensor of an IIoT device (D) is assigned an unique identity $DId =$ ($Id_D || IoT_{MAC}$), where $Id_D$ is the device identifier and $IoT_{MAC}$ is the sensor's MAC address. Our proposed scheme has three major parts which are detailed out below. The scheme is also presented through sequence diagram in Fig. \ref{fig:Image2}.

\noindent \textbf{Part 1 : Initialization}

\noindent In this part, each of the parties involved in the communication converge on the security parameters $<\lambda, \mathbb{G}_1, \mathbb{G}_T, g, q>$ and derive their necessary keys required for computation.

\noindent \textbf{Part 2 : Data Generation to Storage}

\noindent This part is subdivided into (a) Data Generation (b) Data Storage. The first sub-part deals with secure sharing of the generated data from IIoT devices to fog nodes. In the second sub-part, fog nodes process all such received data $m_i$.  F provides a feedback to D for each $m_i$ it receives and then stores the summaries of such data to a file $f$. The file is then encrypted to $f'$ for effectively storing it in cloud, so that it can later be retrieved by the concerned parties. For example, in an industry the periodic temperature readings from sensors placed in the furnace maybe recorded and an hourly average of such values can be stored as a summary to a file $f$. In this stage, apart from encrypting $f$, F defines access policies related to $f$ and also encrypts the key used for file encryption. Lastly, Cl verifies the integrity of the received file.

\begin{algorithm}[!htb]
\small
\DontPrintSemicolon
\SetKwProg{Fn}{Function}{:}{}
\SetKw{KwEnd}{end}
\footnotesize
\SetKwFunction{FAesKey}{Generate\_AesKey}
\Fn{\FAesKey{$1^{\lambda}$}}{
Derive symmetric key $K_F$\;
\KwRet {$K_F$}\;
}

\SetKwFunction{FKeygen}{KeyGenF}
\Fn{\FKeygen{$\lambda$}}{
Retrieve ($params$ , $\mathbf{S}$) from Blockchain B\;
Select a random element $x \in \mathbb{Z}^*_q$.\; Assign $sk_F = x \in \mathbb{Z}^*_q$ and $pk_F = g^{sk_F} \in \mathbb{G}_1$\;
\KwRet {($sk_F$ , $pk_F$)}\;
  }

\SetKwFunction{FAesEncrypt}{AesEncrypt}
\Fn{\FAesEncrypt{$f$}}{
Generate $ciphertext$ ($f'$) over $f$ using $K_F$ and random $nonce$\;
Store $f'$ in Cl and value of $nonce$ in B\;
}

\SetKwFunction{FEncrypt}{Encrypt\_AesKey}
\Fn{\FEncrypt{$K_F$}}{
Retrieve public key of C ($pk_C$) from B\;
Select a random element $r \in \mathbb{Z}^*_q$.\; Compute $c'$ = $<c_1 = ({pk_F})^r$ , $c_2 = \hat{e}(g,g)^r.K_F>$\;
Set access policy ($accessP$) with respect to $c'$ in B\;
Calculate $h_1$ = $H (f')$ and $h_2$ = $H (c')$\;
Store $Meta_1=(DId||h_1||h_2)$ in B\;
\KwRet {($c'$ = $<c_1$ , $c_2>$)}\;
  }

\SetKwFunction{FReEncrypt}{ReEncrypt\_AesKey}
\Fn{\FReEncrypt{$Req$}}{
\uIf{($Req$) received from Cl}{
Compute $rk_{F \rightarrow C} = {pk_C}^{1/{sk_F}}$\;
Select a random element $t \in \mathbb{Z}^*_q$.\; Compute $c''$ = $<c'_1 = {rk^{1/t}_{F \rightarrow C}}$ ; $c''_1 = c^t_1>$\;
Calculate $h_3$ = $H (c'')$. Store $Meta_2 = (DId||h_3)$ in B\;
Lock $P_4$ coins in $\mathbf{S}$. Send $c''$ = $<c'_1$ , $c''_1$ , $c_2>$ to C\;
}
}
\caption{Functions executed by Fog Node (F)}
\label{algo:Algo1}
\end{algorithm}

\begin{algorithm}[!ht]
\small
\DontPrintSemicolon
\SetKwProg{Fn}{Function}{:}{}
\SetKw{KwEnd}{end}
\footnotesize
\SetKwFunction{FVIntegrity}{Verify\_FileIntegrity}
\Fn{\FVIntegrity{$f'$}}{
Calculate $h'_1$ = $H(f')$ and compute $res$ = Verify\_FileHash($h'_1$) \;
\uIf{($res$ == false)}{
\KwRet {Abort}\;
}
\uElse{
\KwRet {$true$}\;
}
}

\SetKwFunction{FVerify}{Verify\_AccessPolicy}
\Fn{\FVerify{$Req$}}{
\uIf{($Req$) received from C}{
Compute $res$ = Compare\_AccessPolicy($Req$)\;
\uIf{($res$ == true)}{
Send $Req$ to F\;
\KwRet {true}\;
}
\uElse{
\KwRet {Abort}\;
}
}
}

\SetKwFunction{FFS}{SendFile}
\Fn{\FFS{}}{
Compute commitment $c$, $z$ from $f'$, $r_z$ and $r_{\phi}$\; \tcp{\footnotesize $z$ = Encoded output; $r_z$ and $r_{\phi}$ are Merkle-tree root elements} 
Send $c$, $r_z$, $r_{\phi}$ to $\mathbf{J}$ and predicate $\phi$, $z$ to C\;
\uIf{(Contract Accepted)}{
Lock $P_2$ coins in $\mathbf{J}$. Send key $k$ to $\mathbf{J}$\;
\uIf{($D$ == "valid")}{
Receive $P_1$ + $P_2$ coins from $\mathbf{J}$\;
\KwRet {}\;
}
\uElseIf{($D$ == "valid complain")}{
$\mathbf{J}$ seizes Cl's deposit (i.e. $P_2$) and returns $P_1$ + $P_2$ coins to C\;
\KwRet {}\;
}
\uElse{
Call $\mathbf{J}$ to trigger release of $P_1$ + $P_2$ coins in its favor\;
\KwRet {}\;
}
}
\uElse{
Terminate the protocol\;
\KwRet {}\;
}
}

\caption{Functions executed by Cloud (Cl)}
\label{algo:Algo2}
\end{algorithm}

\begin{algorithm}[!ht]
\small
\DontPrintSemicolon
\SetKwProg{Fn}{Function}{:}{}
\SetKw{KwEnd}{end}
\footnotesize
\SetKwFunction{FKeygen}{KeyGenC}
\Fn{\FKeygen{$\lambda$}}{
Retrieve ($params$ , $\mathbf{S}$) from Blockchain B\;
Select a random element $y \in \mathbb{Z}^*_q$. Assign $sk_C = y \in \mathbb{Z}^*_q$ and $pk_C = g^{sk_C} \in \mathbb{G}_1$. Store $pk_C$ in B\;
\KwRet {($sk_C$ , $pk_C$)}\;
}

\SetKwFunction{FReq}{SendAceesRequest}
\Fn{\FReq{}}{
Generate $Req$ for accessing file $f'$ and send it to Cl\;
}

\SetKwFunction{FFP}{ReceiveFile}
\Fn{\FFP{}}{
Receive $c$, $r_z$, $r_{\phi}$ from $\mathbf{J}$ and $\phi$, $z$ from Cl\;
Compute $r'_z$ and $r'_{\phi}$ using $\phi$ and $z$\;
\uIf{($r'_z$ == $r_z$ \&\& $r'_{\phi}$ == $r_{\phi}$) }{
Lock $P_1$ coins in $\mathbf{J}$ and send "Contract Accepted" to $\mathbf{J}$\;
\uIf{($k$) not received}{
Cl is \textbf{malicious}. $\mathbf{J}$ sends back $P_1$ + $P_2$\;
Terminate the protocol\;
\KwRet {}\;
}
\uElse{
Compute $f'$ using $z$, $\phi$ and $k$\;
Either claim $f'$ to be valid or generate a complain with a valid proof of misbehavior. Send decision $D$ to $\mathbf{J}$\;
}
}
\uElse{
\KwRet {Abort}\;
}
}

\SetKwFunction{FRAesKey}{Decrypt\_AesKey}
\Fn{\FRAesKey{}}{
Lock $P_3$ coins in $\mathbf{S}$\;
Receive $c''$ = $<c'_1$ , $c''_1$ , $c_2>$ from F.
Calculate $h'_3$ = $H(c'')$. Compare $h'_3$ using function $res$ = Verify\_KeyHash($h'_3$) \;
\uIf{($res$ == false)}{
$\mathbf{S}$ seizes F's deposit (i.e. $P_4$) and C receives $P_3+P_4$ coins\;
\KwRet {Abort}\;
}
\uElse{
$\mathbf{S}$ returns $P_3+P_4$ coins to F\;
Calculate $K_F$ = $c_2/\hat{e}(c'_1,c''_1)^{1/sk_C}$\;
\KwRet {$K_F$}\;
}
}

\SetKwFunction{FDecrypt}{DecryptFile}
\Fn{\FDecrypt{$f'$}}{
Read $nonce$ value from B\;
Derive file $f$ from $f'$ using $K_F$ and $nonce$\;
\KwRet {$f$}\;
}
\caption{Functions executed by Client (C)}
\label{algo:Algo3}
\end{algorithm}

\begin{algorithm}[!ht]
\small
\DontPrintSemicolon
\SetKwProg{Fn}{Function}{:}{}
\SetKw{KwEnd}{end}
\footnotesize
\SetKwFunction{FRegP}{RegisterParams}
\Fn{\FRegP{}}{
A trusted authority deposits global parameters $params$ $<\lambda, \mathbb{G}_1, \mathbb{G}_T, g, q>$ and $\mathbf{S}$\;
\KwRet {($contract-address$)}\;
}

\SetKwFunction{FRegF}{RegisterFog}
\Fn{\FRegF{$pk_F$, $contractTerms$}}{
F deposits required money as per $contractTerms$\;
\KwRet {$FogID$}\;
}

\SetKwFunction{FRegC}{RegisterCloud}
\Fn{\FRegC{$contractTerms$}}{
Cl deposits required money as per $contractTerms$\;
\KwRet {$CloudID$}\;
}

\SetKwFunction{FRegCt}{RegisterClient}
\Fn{\FRegCt{$pk_C$, $contractTerms$}}{
C deposits required money as per $contractTerms$\;
\KwRet {$ClientID$}\;
}

\SetKwFunction{FVf}{Verify\_FileHash}
\Fn{\FVf{$h'$}}{
Compare received $h'$ with stored $h$ for file $f'$\;
\uIf{($h'$ == $h$)}{
\KwRet {$true$}\;
}
\uElse{
\KwRet {$false$}\;
}
}

\SetKwFunction{FCap}{Compare\_AccessPolicy}
\Fn{\FCap{$Req$}}{
Compare received $Req$ with stored $\rho$ for file $f$\;
\uIf{(Verified successfully)}{
\KwRet {$true$}\;
}
\uElse{
\KwRet {$false$}\;
}
}

\SetKwFunction{FVc}{Verify\_KeyHash}
\Fn{\FVc{$h'$}}{
Compare received $h'$ with stored $h$ for $c''$\;
\uIf{($h'$ == $h$)}{
\KwRet {$true$}\;
}
\uElse{
\KwRet {$false$}\;
}
}
\caption{Controller Contract ($\mathbf{S}$) Functions}
\label{algo:Algo4}
\end{algorithm}

\noindent \textbf{Part 3: Data Request to Retrieval}

\noindent This part is subdivided into (a) Data Request (b) Data Retrieval. The execution is triggered when C sends a request $Req$ to access $f$ along with an incentive to Cl. Cloud then processes $Req$ and depending on whether C has access rights to $f$, data retrieval begins. Here, Cl securely shares $f'$ and F shares the key required to decrypt $f'$ with C.

\begin{figure}[!ht]
\begin{center}  
\fbox{\small
\begin{minipage}{\columnwidth}
The scheme describes behaviours of honest Fog node $\mathbf{F}$, Cloud $\mathbf{Cl}$ and Client $\mathbf{C}$\\
\underline{\textbf{Initialization:}}\\
A trusted authority computes $params$ $<\lambda, \mathbb{G}_1, \mathbb{G}_T, g, q>$ and registers them in Controller Contract $\mathbf{S}$ by calling $RegisterParams()$.\\
$\mathbf{F:}$ Compute symmetric key $K_F=Generate\_AesKey(1^{\lambda})$ and key-pair $(sk_F,pk_F) \leftarrow KeyGenF(\lambda)$.\\ Register $\mathrm{FogID}=RegisterFog(pk_F,contractTerms)$ in $\mathbf{S}$.\\
$\mathbf{C:}$ Compute key-pair $(sk_C,pk_C) \leftarrow KeyGenC(\lambda)$. \\Register $\mathrm{ClientID}=RegisterClient(pk_C,contractTerms)$ in $\mathbf{S}$.\\
$\mathbf{Cl:}$ Register $\mathrm{CloudID}=RegisterCloud(contractTerms)$ in $\mathbf{S}$.\\
\underline{\textbf{Data Generation to Storage:}}\\
$\mathbf{F:}$ Process and store summary of each received data $m_i$ to a file $f$.\\ Compute encrypted $f'= AesEncrypt(f)$ and store $f'$ in $\mathbf{Cl}$. \\Compute encrypted $K_F$ as $ (c'=<c_1,c_2>) \leftarrow$ $Encrypt\_AesKey(K_F)$. \\Store access policy $(\rho)$ and ${Meta}_1$ in contract $\mathbf{S}$.\\
$\mathbf{Cl:}$ On receiving $f'$, execute $Verify\_FileIntegrity(f')$, where $\mathbf{S}$ returns $\textit{true} \leftarrow Verify\_FileHash(H(f'))$ only for correct a $f'$.\\
\underline{\textbf{Data Request to Retrieval:}}\\
$\mathbf{C:}$ Generate $\mathrm{Req}=SendAccessRequest()$ and forward it to $\mathbf{Cl}$.\\
$\mathbf{Cl:}$ On receiving $\mathrm{Req}$, execute $Verify\_AccessPolicy(\mathrm{Req})$ where $\mathbf{S}$ returns $\textit{true} \leftarrow Compare\_AccessPolicy(\mathrm{Req})$, only for an authorized $\mathbf{C}$. After receiving $\textit{true}$ from $\mathbf{S}$, $\mathbf{Cl}$ initiates fair exchange of encrypted file $f'$ with $\mathbf{C}$ using $SendFile()$ and concurently intimidates $\mathbf{F}$ about $\mathrm{Req}$.\\
$\mathbf{C:}$ Trigger $ReceiveFile()$ to securely receive encrypted file $f'$ from $\mathbf{Cl}$ where payments are handled by Judge Contract $\mathbf{J}$.\\
$\mathbf{F:}$ On receiving $\mathrm{Req}$, compute re-encrypted ciphertext $(c''=<c'_1,c''_1,c_2>) \leftarrow$ $ReEncrypt\_AesKey(\mathrm{Req})$ and send $c''$ to $\mathbf{C}$.\\
$\mathbf{C:}$ On receiving $c''$ from $\mathbf{F}$, execute $Decrypt\_AesKey()$ which derives $K_F$ after $\mathbf{S}$ returns $\textit{true} \leftarrow Verify\_KeyHash(H(c''))$, for a correct $c''$. Finally, recover file $f=DecryptFile(f')$.
\end{minipage}
}
\caption{\small \sl Execution of \textit{FairShare} for honest participants \label{fig:Image8}}
\end{center}  
\end{figure}

\noindent \textbf{Algorithmic Construction:} To simplify the algorithm design, we have made the following assumptions :

\begin{itemize}
    \item We assume a single fog and single client scenario.
    \item The basic interactions between D and F for the purpose of data generation are omitted here, as they do not involve any on-chain communications.
    \item We also assume that a single summarized file $f$ is generated and exchanged between the concerned parties.
\end{itemize}

All these assumptions can be extended to obtain a multi-client generic setting, which has been omitted to attain simplicity. The functions executed by fog, cloud, client and controller contract are described in Procedures \ref{algo:Algo1}, \ref{algo:Algo2}, \ref{algo:Algo3} and \ref{algo:Algo4} respectively. Preciously, Fig. \ref{fig:Image8} shows execution of \textit{FairShare} for honest participants. For a more detailed step-wise description of \textit{FairShare}, interested readers can refer to \textit{Appendix \ref{AppC}}.

\noindent It is evident from the above discussion that to share file $f$ with multiple recipients (i.e. $C_1$, $C_2$, \ldots ,$C_n$), F just needs to update access policy ($\rho$) and generate re-encryption keys for each individual recipient. F will then perform re-encryption on the shared key ($K_F$) and send it to the respective Client for which it gets rightfully paid. Thus, this is unlike the traditional data sharing schemes where F would have to encrypt file $f$ individually for sharing it with each recipient $C_1$, $C_2$, \ldots ,$C_n$ thereby increasing computation overhead. Moreover, for user revocation, F only needs to update access policy ($\rho$) in controller contract $\mathbf{S}$. Thus, user management in \textit{FairShare} is less cumbersome and quite computationally less intensive compared to the existing data sharing schemes. 

\section{Security Analysis \label{Section5}}

\noindent \textbf{Theorem:} \textit{If the hash function $H$ is collusion-resistant, the encryption as well as re-encryption algorithms are secure and blockchain $B$ is resilient to tampering, then an adversary controlling at most two of the three parties, cannot compromise authenticity, privacy, accountability and fairness, except with negligible probability.}

\noindent \textbf{Proof:} We break this proof into six cases, each representing a combination of malicious entities amongst Fog node (F), Cloud (Cl) and Client (C). We omit the cases where all the parties are honest as well as where all the three parties are malicious. This is because in the first case the scheme would terminate safely and for the second case we have already described that at least one of the three parties should be honest for a fair execution of the proposed scheme.

\noindent\textbf{Case I: Malicious Fog Node}

The Fog Node can behave maliciously while sharing $c''$ in \textbf{Part 3} with Client, by sending a different $\hat{c''}$ to C despite computing the correct $c''$. However, according to \textit{FairShare}, F calls function $ReEncrypt\_AesKey()$ which signs and sends hash of $c''$ as a transaction to blockchain B before sharing $\hat{c''}$ with C. Now, when C receives an incorrect $\hat{c''}$ from F, it proceeds to call function $Decrypt\_AesKey()$ which verifies hash of received $\hat{c''}$ with the hash stored previously by F in B by calling contract $\mathbf{S}$. For a malicious F, this verification fails and F gets penalized as per agreements made in $\mathbf{S}$, thereby ensuring accountability and fairness. 

It can also happen that F generates the correct $f'$ from $f$ in \textbf{Part 2}, however while storing $f'$ in Cloud, it supplies a different $\hat{f'}$. But, we have already discussed that players do not deviate from the scheme without sufficient incentive and here, F doesn't have any real incentive to do so. Hence, the question of authenticity breach of the file doesn't arise. Finally, privacy property is satisfied as Cloud is honest, hence it doesn't attempt to extract any information about $f$/$f'$ and $\rho$ from F.

\noindent\textbf{Case II: Malicious Cloud}

A malicious cloud may attempt to compromise privacy of the stored data by trying to extract information about $f'$ using its corresponding access policy $\rho$. However, file $f'$ is encrypted with a key known only to F and later shared with C. Since both the parties (F and C) are honest, therefore privacy of $f'$ is preserved. A malicious cloud may also try to attack the authenticity property by claiming to have received a different $\hat{f'}$ ($\neq f'$) from F in \textbf{Part 2}. This may happen either due to errors in communication or due to malicious intent of Cl. Correspondingly, Cl may also forge a correctly received $f'$ to $\hat{f'}$ while sharing it with C to initiate another attack on the authenticity property in \textbf{Part 3}. However, in the first case F calls function $Encrypt\_AesKey()$ which signs and sends hash of $f'$ as a transaction to B before sharing $f'$ with Cl. Now, when Cl receives $f'$ and claims it to be $\hat{f'}$, Cl has to call smart function $Verify\_FileIntegrity(\hat{f'})$ to verify hash of the forged $\hat{f'}$. This verification naturally fails and the process gets \textit{aborted}, thus ensuring authenticity of $f'$. Similarly, in the second case, as C is honest therefore on receiving an incorrect $\hat{f'}$, it will lodge a complain with Judge contract ($\mathbf{J}$) along with a valid proof of misbehavior.  $\mathbf{J}$ will then verify the complain and penalize Cl for behaving incoherently. Thus, authenticity, accountability and fairness are all preserved as C couldn't be deceived with a forged $\hat{f'}$ and Cl gets rightfully penalized for behaving maliciously.

\noindent\textbf{Case III: Malicious Client}

A Client (C) might behave maliciously by requesting data $f'$ to Cl for which s/he is not authorized. Since, F stores access policy ($\rho$) respective to each file in $\mathbf{S}$ so, whenever C sends such a request, Cl calls $Verify\_AccessPolicy()$ to find whether C is authorised to view $f'$. This in turn triggers $\mathbf{S}$ to verify the request, where $\mathbf{S}$ detects the misbehavior and safely aborts the process, thus maintaining privacy of the stored data.

The client might also behave maliciously after receiving $f'$ by complaining that it has received an incorrect $\hat{f'}(\neq f')$ and is therefore unable to retrieve $f$ from it. However, this condition is impossible to arise, because according to the agreement made in $\mathbf{J}$, C has to generate a valid proof of misbehavior against Cl, if C has to claim for an incorrect data. To do so, C needs to find a collision in the hash function which we have assumed to be collision resistant. Thus, our proposed scheme is secure against such attacks.

Finally, accountability, fairness and authenticity are guaranteed in this case as F and Cl are both honest.

\noindent\textbf{Case IV: Malicious Fog Node and Cloud}

If both the parties become malicious, they can collude when F sends an incorrect $f'$ and Cl agrees to it. However, the request for verifying integrity of the incorrect $\hat{f'}$ is handled by controller contract function $Verify\_FileHash()$. Since, a smart contract cannot be tampered, therefore it is practically infeasible to launch such an attack, as $\mathbf{S}$ will safely \textit{abort} the scheme on failure of the verification. Therefore, the case reduces to that a malicious cloud, and hence authenticity, accountability and fairness are all preserved.

The adversary however has access to the file because it controls both the fog node and the cloud. Hence, in this case the privacy property is not applicable. 

\noindent\textbf{Case V: Malicious Fog Node and Client}

In \textbf{Part 2}, there is an exchange of $nonce$ and $pk_C$ between these two parties. However, these parameters are sent through $\mathbf{S}$ and hence no attacks are possible. 

Given the above condition holds, these two parties may collude when F sends an incorrect $\hat{c''}(\neq c'')$ to C yet C doesn't complain. However, verification of received $\hat{c''}$ is carried out by controller contract function $Verify\_KeyHash(\hat{c''})$. So, for an incorrect $\hat{c''}$ the verification fails, and $\mathbf{S}$ safely aborts the process. Thus, such a colluding scenario will never arise in reality and hence the case reduces to that of a malicious F only. Thus, we can claim that authenticity, accountability and fairness are preserved, as $\mathbf{S}$ safely \textit{aborts} the process and handles the necessary monetary transactions. Lastly, privacy is also preserved since Cl is honest.

\noindent\textbf{Case VI: Malicious Cloud and Client}

If both these parties become malicious, they can collude on two cases. In the first case, when C requests to access a file for which s/he is not authorized and Cl grants the request. However, as already explained in $\textit{Case III}$, this comparison is carried out by $\mathbf{S}$ (function $Compare\_AccessPolicy()$) which safely \textit{aborts} the access request for an invalid C. Thus, this colluding scenario is impractical to arise. The second case arises when Cl sends an incorrect $\hat{f'}(\neq f')$ yet C doesn't complain, neither does it generate a proof of misbehavior against Cl as per function $ReceiveFile()$. Ideally, C proceeds to execute function $DecryptFile()$, but becomes unable to derive the requested file $f$ due to an incorrect input $\hat{f'}$. Therefore, we can easily conclude that C doesn't have any real motive to collude with a malicious Cl as it is unable to achieve any extra benefit from such a collusion. Henceforth, this case reduces to that of a malicious cloud only. Thus, we can claim that authenticity, privacy, accountability and fairness are all preserved in this case.

Apart from the scenarios discussed above, we have considered all other possibilities and observed that such situations will never arise in reality. For example, client can try to behave innocently by claiming to have not received $c''$, even after receiving it correctly. However, as discussed players do not deviate from the scheme without sufficient incentive. Here, C has already deposited $P3$ coins in the contract $\mathbf{S}$ and it also wants the final file $f$, therefore it doesn't have any real incentive to act maliciously.

\begin{table}[!htb]
\centering
\caption{\small \sl A comparative summary of key features}
\label{tab:Table5}
\scalebox{0.68}{%
\begin{tabular}{c |c|c|c|c}
\hline
\textbf{Features} & \textbf{BPREET \cite{r23}} & \textbf{Work \cite{r26}} & \textbf{IBPRE Scheme for PEC \cite{r25}} & \textbf{\textit{FairShare}} \\ \hline
\textbf{Correctness} & $\checkmark$ & $\times$ & $\checkmark$ & $\checkmark$ \\ \hline
\textbf{Confidentiality} & $\checkmark$ & $\checkmark$ & $\checkmark$ & $\checkmark$ \\ \hline
\textbf{Decentralization} & $\checkmark$ & $\checkmark$ & $\checkmark$ & $\checkmark$ \\ \hline
\textbf{Authenticity} & $\times$ & $\checkmark$ & $\checkmark$ & $\checkmark$ \\ \hline
\textbf{Integrity} & $\times$ & $\checkmark$ & $\checkmark$ & $\checkmark$ \\ \hline
\textbf{Privacy} & $\times$ & $\times$ & $\times$ & $\checkmark$ \\ \hline
\textbf{Accountability \& Fairness} & $\times$ & $\times$ & $\times$ & $\checkmark$ \\ \hline
\end{tabular}%
}
\end{table}

Table \ref{tab:Table5} shows a comparative summary of \textit{FairShare} with three state-of-the art papers \cite{r23,r25,r26} on the basis of key security features achieved by these schemes. It is evident that \textit{FairShare} outperforms the other schemes considerably.

\section{Performance Evaluation \label{Section6}}

In this section, we evaluate the performance of \textit{FairShare} both theoretically and experimentally.

\subsection{Theoretical Analysis}

The computation and communication overheads are measured in terms of execution time and number of transmitting bytes respectively. During analysis, we consider Type A pairings where each group element is of size 128 bytes \cite{i2}. We consider Advanced Encryption Standard (AES) in GCM-256 and Rivest, Shamir, Adleman (RSA) for symmetric and asymmetric encryption respectively. However, they could also be replaced with any other optimized and lightweight encryption algorithms. The corresponding ciphertext lengths of AES and RSA are 16 and 256 bytes respectively. Table \ref{tab:Table2} summarizes the notations used for theoretical analysis.

\begin{table}[!htb]
\centering
\caption{\small \sl Notations used for Theoretical Analysis}
\label{tab:Table2}
\scalebox{0.80}{
\begin{tabular}{c|c}
\hline
\rowcolor[HTML]{9B9B9B} 
\textbf{Notations} & \textbf{Meaning} \\ \hline
$T_H$ & Time taken to perform a hash operation \\ \hline
$T_{SG}$ & Time taken to generate a signature \\ \hline
$T_{XOR}$ & Time taken to perform a XOR operation \\ \hline
$T_V$ & Time taken to verify a signature \\ \hline
$T_E$ & Time taken to perform an exponentiation \\ \hline
$T_P$ & Time taken to perform a pairing operation \\ \hline
$T_M$ & Time taken to perform a multiplication \\ \hline
$T_{KS}$ & Time taken to generate a symmetric key \\ \hline
$T_{KA}$ & Time taken to generate an asymmetric key-pair \\ \hline
$T_{PRE}$ & Time taken to generate a re-encryption key-pair \\ \hline
$T_{SE}/T_{SD}$ & Time taken to perform one symmetric encryption/decryption \\ \hline
$T_{AE}/T_{AD}$ & Time taken to perform one asymmetric encryption/decryption \\ \hline
|Req| & Size of one request message \\ \hline
|f| & Size of a file $f$ \\ \hline
\end{tabular}%
}
\end{table}

\noindent The overheads are calculated for the processing and exchange of a single file $f$. We observe (Table \ref{tab:Table4}) that the computation overheads for Data Request and Data Retrieval are comparatively less than Data Generation and Data Storage. This is because Data Generation has two hash, a signature generation/verification, three XOR and several encryption/decryption operations while Data Storage has two exponentiations, three XOR, one hashing, one pairing and one encryption operation which are obviously more computationally intensive compared to the tasks (e.g. single hash or XOR or division) performed during Data Request and Retrieval. This justifies our proposal because the same file $f$ maybe requested by multiple clients and thus having lesser data retrieval time is beneficial. Finally, the communication overhead for Data Retrieval is slightly higher than Data Generation because of the additional security provided while sharing the requested file with the Client.

\subsection{Experimental Analysis through Simulation}

Here, we implement and analyze a prototype of \textit{FairShare}.

\noindent\textbf{Setup:} We have used Ethereum \cite{c8} as the blockchain platform to implement a prototype and evaluate our proposed scheme. Our entire code is approximately 2000 lines, consisting of Go-Ethereum\footnote[4]{\url{https://github.com/ethereum/go-ethereum}} (the most popular Ethereum implementation) modifications written in Golang and smart contracts written in Solidity\footnote[5]{\url{https://solidity.readthedocs.io/en/v0.6.10/}}. To perform  arithmetic in $\mathbb{Z}_p$, bilinear pairing computations and elliptic curve group related operations, we have used the Golang wrapper\footnote[2]{\url{https://godoc.org/github.com/Nik-U/pbc}} of the popular \textit{PBC Library} \cite{l5}. We have used Type A pairings based on the elliptic curve, $y^2=x^3+x$. For the FairSwap \cite{c3} implementation, we have referred to their smart contract pseudo code to develop our own Golang code and Solidity bindings.

We have deployed our implementation\footnote[3]{\url{https://tinyurl.com/FairShareCodes}} on a private Ethereum network consisting of two nodes as well as on test network. We have used a single machine with Intel® Core™ i5-7200U CPU @ 2.50GHz and 8GB of RAM running 64-bit Ubuntu 18.04.3 LTS. The controller contract code is available at address 0x3384922dF8f0fbce24D7F60A843FE5b09f9eD1a2 in Ropsten testnet of Ethereum. Fog node, cloud and client are running alongside the Ethereum nodes.

We know from existing literature \cite{c12} that amount of data generated per hour from a sensor for a typical industrial environment is approximately 14 MB. We also know that redundant data from the sensor can then be reduced by around 65\% at fog nodes \cite{c13} which brings down the data size to approximately 5MB for an hour. Further, as per our proposal the fog nodes process these raw data and store a summarized version of such data to cloud in the form of a file. Hence, we have chosen different file sizes ranging from 1KB to 10MB for our evaluation considering that summarized data for a shorter time frame (say 2-3 hours) are stored in a file and uploaded to cloud for better scheduling, maintenance and diagnosis.

\noindent\textbf{Evaluation Metrics: \label{em}} We use three metrics to evaluate our prototype such as computation time, blockchain execution time and overall execution time. Computation time refers to the time taken by the individual parties (Fog, Cloud, Client) to perform various operations (e.g. encryption). Blockchain execution time is the summation of the blockchain call time and the time taken by the smart contract to perform certain operations (e.g. integrity verification). Here, blockchain call time means the time taken to mine a transaction into blockchain or read a data from the blockchain. Finally, the overall execution time is the summation of the computation time and the blockchain execution time.

\begin{table}[!t]
\centering
\caption{\small \sl Theoretical Overhead Analysis of \textit{FairShare}}
\label{tab:Table4}
\scalebox{0.75}{
\begin{tabular}{c|c|c}
\hline
\rowcolor[HTML]{9B9B9B} 
Tasks & \begin{tabular}[c]{@{}c@{}}Computation \\ Overhead\end{tabular} & \begin{tabular}[c]{@{}c@{}}Communication\\ Overhead (bytes)\end{tabular} \\ \hline
Initialization & 2($T_{KS}$ + $T_{KA}$ + $T_{PRE}$) & - \\ \hline
Data Generation & \begin{tabular}[c]{@{}c@{}}2$T_H$ + $T_{SG}$ + 3$T_{XOR}$ + $T_{SE}$\\ + $T_{AE}$ + $T_{AD}$ + $T_{SD}$ + $T_V$\end{tabular} & 272 \\ \hline
Data Storage & \begin{tabular}[c]{@{}c@{}}$T_{SE}$ + 2$T_E$ + $T_P$ \\ + $T_M$ + 2$T_{XOR}$ + 2$T_H$\end{tabular} & |f| \\ \hline
Data Request & 3$T_E$ + $T_H$ + $T_{XOR}$ & |Req| \\ \hline
Data Retrieval & $T_H$ + $T_D$ + $T_P$ + $T_E$ +$T_{SD}$ & |f| + 384 \\ \hline
\end{tabular}%
}
\end{table}

\noindent\textbf{Results and Discussion:} The primary objective of our prototype implementation is to observe the overhead introduced by blockchain. We conduct five sets of experiments. For each set, the average results of ten independent runs respective to each parameter/input file size has been registered. Among these, in the first three set of experiments we compare the performance of \textit{FairShare} with a state-of-the art IBPRE scheme for Pervasive Edge Computing (PEC) \cite{r25} where the same environment has been used as that of \textit{FairShare}. We chose the work \cite{r25} for comparison because it is the closest competitor to \textit{FairShare}. However, the rest of the experiments deal only with the performance evaluation of \textit{FairShare} because it is evident from Table \ref{tab:Table5} that the work \cite{r25} doesn't address the major security features required for an industrial IoT scenario, hence comparing it with \textit{FairShare} is irrelevant.

\begin{figure}[!htb]
    \begin{minipage}[t]{.45\linewidth}
        \centering
        \includegraphics[scale=0.22]{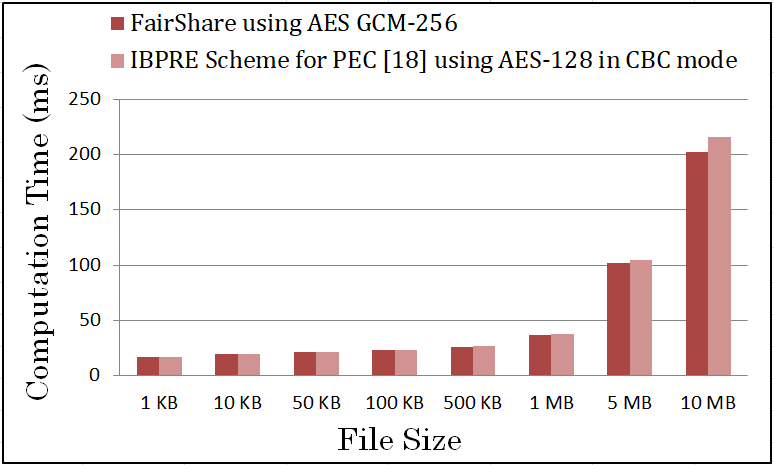}
        \subcaption{Encryption by Fog node}\label{fig:Image4a}
    \end{minipage}
   \hspace{0.3cm}
    \begin{minipage}[t]{.45\linewidth}
        \centering
        \includegraphics[scale=0.22]{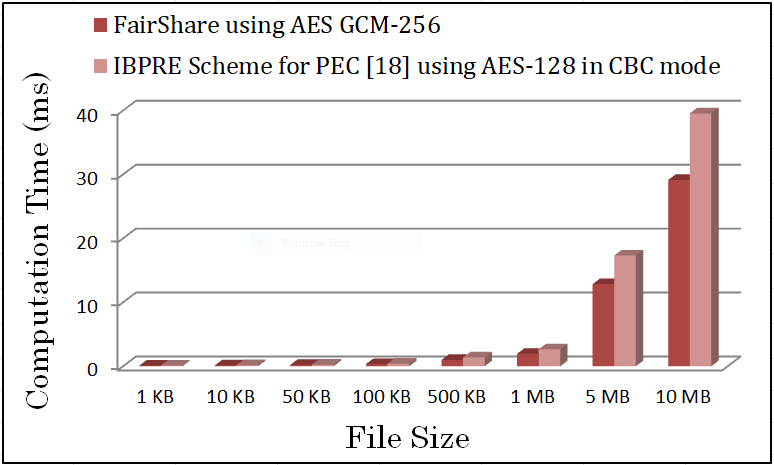}
        \subcaption{Decryption by Client}\label{fig:Image4b}
    \end{minipage}
    \caption{\small \sl Comparison of Computation Time for different $|F|$ using AES} 
    \label{fig:Image4}
    \end{figure}

\begin{figure*}[!htb]
\begin{minipage}[b]{0.26\linewidth}
\centering
\includegraphics[width=0.98\textwidth, height=1.1in]{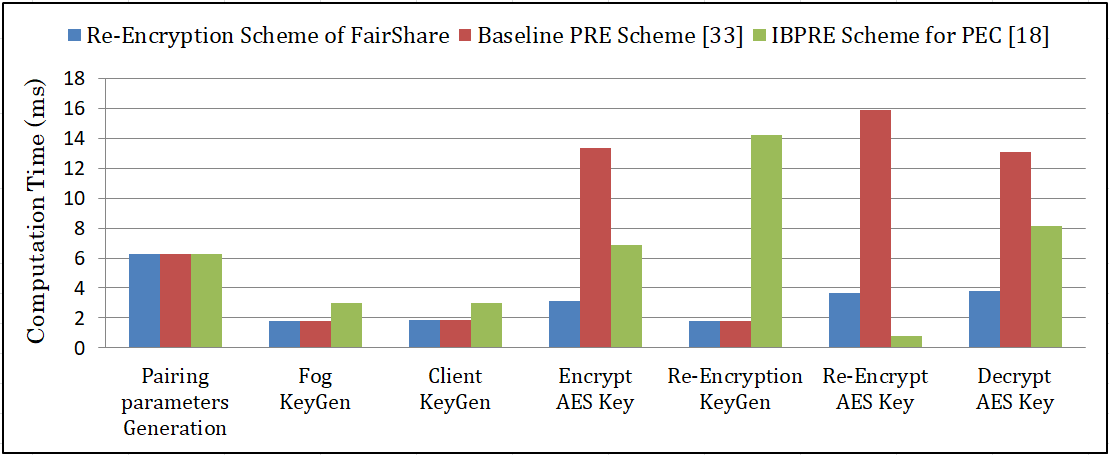}
\caption{\small \sl Comparison of Computation Time of various tasks in Re-Encryption}
\label{fig:Image3}
\end{minipage}
\hspace{0.2cm}
\begin{minipage}[b]{0.22\linewidth}
\centering
\includegraphics[width=0.99\textwidth, height=1.1in]{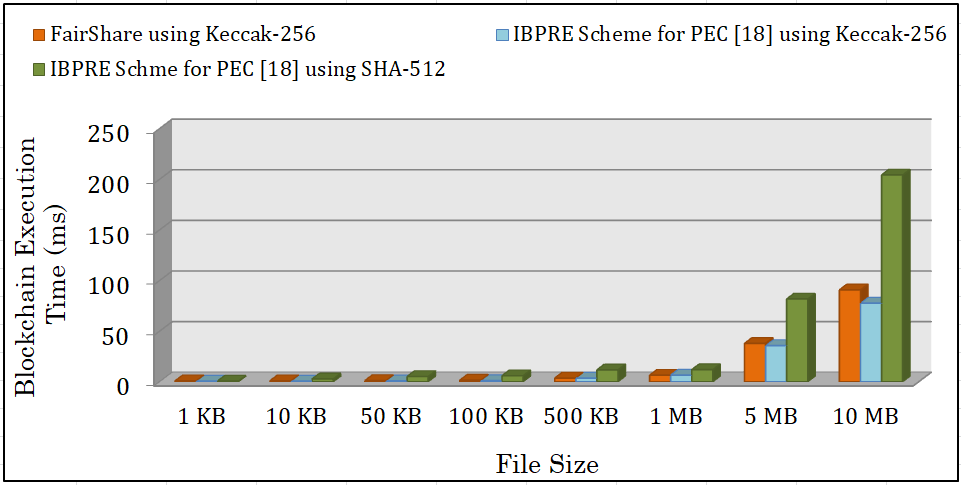}
\caption{\small \sl Blockchain Execution Time for File Integrity Verification with varying $|F|$}
\label{fig:Image5}
\end{minipage}
\hspace{0.5cm}
\begin{minipage}[b]{0.21\linewidth}
\centering
\includegraphics[width=\textwidth, height=1.22in]{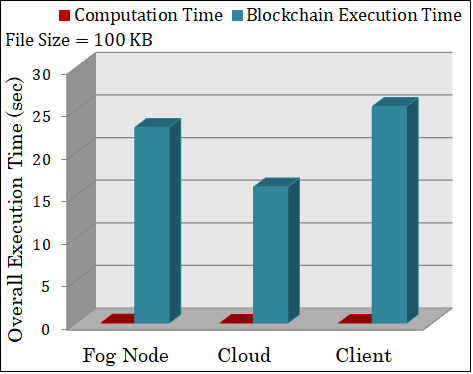}
\caption{\small \sl Overall Execution Time for all Parties} 
\label{fig:Image7}
\end{minipage}
\hspace{0.3cm}
\begin{minipage}[b]{0.21\linewidth}
\centering
\includegraphics[width=0.98\textwidth, height=0.92in]{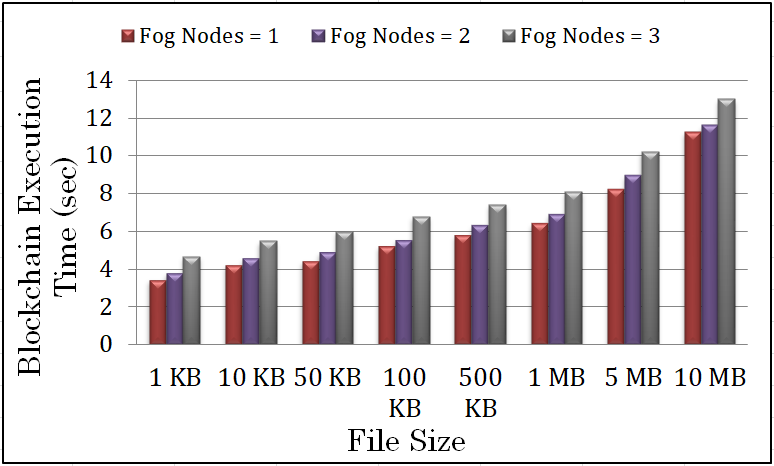}
\caption{\small \sl Blockchain Execution Time for storing metadata about a file in $\mathbf{S}$ with varying $|F|$}
\label{fig:Image10}
\end{minipage}
\end{figure*}

In the first set of experiment, we measure the computation time required by Fog and Client to encrypt/decrypt a file $f$/$f'$ respectively. We plot AES Encryption time (Fig. \ref{fig:Image4a}) and AES Decrytion time (Fig. \ref{fig:Image4b}) with varying file sizes for both the schemes. In our setup, file encryption is performed by Fog node for storing data securely in Cloud whereas file decryption is performed by Client for retrieving the requested data. We observe from Fig. \ref{fig:Image4a}, that encryption time varies negligibly till 1 MB file size, after which it starts growing significantly for larger file sizes. Similarly, we observe from Fig. \ref{fig:Image4b} that decryption time is very small (represented as flat boxes in the figure) till 100KB file size, after which there is a slight linear growth till 1MB, followed by an exponential growth for larger file sizes. This is because AES operates in blocks of fixed size, where the output of each block is fed as an input to the next block (except the first block). For this, each block has to wait for a specific amount of time, before it can begin its processing. Moreover, as the x-axis values in our experimentation are not linear, therefore the computation time growth looks exponential for larger file sizes. Lastly, \textit{FairShare} performs relatively better for both encryption and decryption because AES in GCM mode is parallelizable and therefore is more versatile and faster compared to AES in CBC mode. 

In the second set of experiment, we implement our entire prototype thrice using the modified re-encryption scheme of \textit{FairShare} along with the baseline Proxy Re-Encryption (PRE) scheme \cite{c2} and state-of-the art competitor scheme \cite{r25} separately. We then plot (Fig. \ref{fig:Image3}) computation times of various tasks involved in re-encryption for these prototype implementations. From the figure it is evident that \textit{FairShare}'s re-encryption scheme performs better at most of the tasks (e.g. Encrypt AES Key) while maintaining at per results for the other tasks (e.g. Fog KeyGen). This is because we have suitably modified the baseline scheme \cite{c2} by eliminating certain validity checking conditions at each encryption stage. These validity checks are delegated to blockchain which results in comparably better performance for \textit{FairShare}. The task Re-Encrypt AES Key for the competitor scheme \cite{r25} outperforms \textit{FairShare}, however the cumulative time for Re-Encryption KeyGen and Re-Encrypt AES Key for \textit{FairShare} is much less than that of its competitor.

In the third set of experiment, we plot (Fig. \ref{fig:Image5}) time taken by contract $\mathbf{S}$ to verify integrity of the received file $f'$, for varying file sizes. From the figure, we observe that blockchain execution time for this experiment is nearly constant till 100KB file size, after which there is a linear growth till 1MB, followed by an exponential growth. As explained, the x-axis values in our experimentation are not linear, therefore the execution time growth looks exponential for larger file sizes. We execute the competitor scheme using both SHA-512 (which has been used by \cite{r25}) and Keccak-256 (to show correspondence with our results) hash functions. We observe that, while using Keccak-256, the competitor scheme \cite{r25} has less overheads compared to \textit{FairShare}. This is because, unlike \textit{FairShare} the file integrity verification in \cite{r25} is not executed by contract $\mathbf{S}$. They only retrieve hash value of the file from smart contract and use it for comparison, which doesn't necessarily guarantee authenticity and are prone to other attacks as well (e.g. tampering). Contrarily, while using SHA-512, the overheads for \cite{r25} are more compared to \textit{FairShare}, because of the inherent overhead of the hash function used. Moreover, smart contract related security issues discussed above persists.

\begin{savenotes}
\begin{table}[!t]
\centering
\caption{\small \sl Overheads related to different Smart Contract operations}
\label{tab:Table3}
\scalebox{0.70}{
\begin{tabular}{l |c |c |c |c |c }
\hline
\multicolumn{1}{c|}{\textbf{Tasks}} & {\textbf{\begin{tabular}[c]{@{}c@{}}Execution\\ Time (sec)\end{tabular}}} & {\textbf{\begin{tabular}[c]{@{}c@{}}Gas Usage \\\end{tabular}}} & {\textbf{\begin{tabular}[c]{@{}c@{}}Deployment\\ Cost (ETH)\end{tabular}}} & {\textbf{\begin{tabular}[c]{@{}c@{}}Deployment\\ Cost (USD)\end{tabular}}} & \multicolumn{1}{l}{\textbf{\begin{tabular}[c]{@{}l@{}}Communication\\ Overhead (bytes)\end{tabular}}} \\ \hline
{\begin{tabular}[c]{@{}l@{}}Controller Contract\\ $\mathbf{S}$ Deployment\end{tabular}} & {6.215} & {47002} & {0.001645070} & {4.68} & {-} \\ \hline
{\begin{tabular}[c]{@{}l@{}}Judge Contract\\ $\mathbf{J}$ Deployment\end{tabular}} & {5.080} & {54132} & {0.001894620} & {5.39} & {-} \\ \hline
{Initialization} & {13.18} & {15323} & {0.000536305} & {1.52} & {754} \\ \hline
{\begin{tabular}[c]{@{}l@{}}Data Generation\\ to Storage\end{tabular}} & {\begin{tabular}[c]{@{}c@{}}11.576-\\ 19.525 \footnote[1]{Execution Time is given as a range as it varies with input file size}\end{tabular}} & {3759} & {0.0001315696} & {0.37} & {|F| + 86} \\ \hline
{\begin{tabular}[c]{@{}l@{}}Data Request\\ to Retrieval\end{tabular}} &
{\begin{tabular}[c]{@{}c@{}}33.058-\\ 64.450 * \end{tabular}} &
{6401} & {0.00022404} & {0.64} & {|F| + 170} \\ \hline
\end{tabular}%
}
\end{table}
\end{savenotes}

In the fourth set of experiment, we plot (Fig. \ref{fig:Image7}) the breakup of the overall execution time of \textit{FairShare} for each party considering a fixed file size of 100KB. From the figure, we observe that overall execution time of Client is $\approx 40\%$ (the highest) and that of Cloud is $\approx 24\%$ (the lowest) of the cumulative execution times of all three parties. This proves that we have successfully reduced the trust dependence on cloud and secured the data sharing process by suitably delegating tasks to fog and client. Additionally, for each party, blockchain execution time takes roughly $\approx 99.8\%$ whereas computation time is as low as $\approx 0.2\%$ (represented as flat red boxes) of the overall execution time. This is simply because blockchain related operations are computationally more intensive compared to other tasks due to the inherent overhead of blockchain.

Finally, in the last set of experiment, we show different overheads related to various smart contract operations in Table \ref{tab:Table3}, where |F| is the size of encrypted file $f'$ which is equal to size of $f$. As on 25/04/2022, the standard gas price per unit was $35\ gwei$ using which the corresponding gas price in USD was calculated at an exchange rate of 1ETH = 2842.52 USD. From the table, we observe that the computationally intensive tasks in terms of cost (e.g. contract deployment) are one-time tasks performed during initialization of the system. Therefore, such tasks do not affect the performance of the system in the long run. Thus, we can conclude that both the deployment cost and communication overhead is the resultant summation of costs incurred during Data Generation to Storage and Data Request to Retrieval phases. This means deployment cost is (0.37 + 0.64) = 1.01 USD and communication overhead is (86 + 170) = 256 bytes. We also observe that Data Generation to Storage is the least expensive which is extremely suitable for industrial applications, since huge amount of data need to be processed regularly. The little extra overhead in Data Request to Retrieval part compared to Data Generation to Storage is justified at the cost of achieving accountability and fairness while still guaranteeing fair exchange of file. Thus, \textit{FairShare} is beneficial for industrial environments where data access requests are generated in non-real time to make strategic decisions for improving overall functionality of the IIoT system.

\subsection{Testbed Implementation}

We implement \textit{FairShare} and validate its performance using a larger and a more realistic IIoT testbed setup.

\begin{figure}[!htb]
\begin{center}  
\includegraphics[scale=0.28]{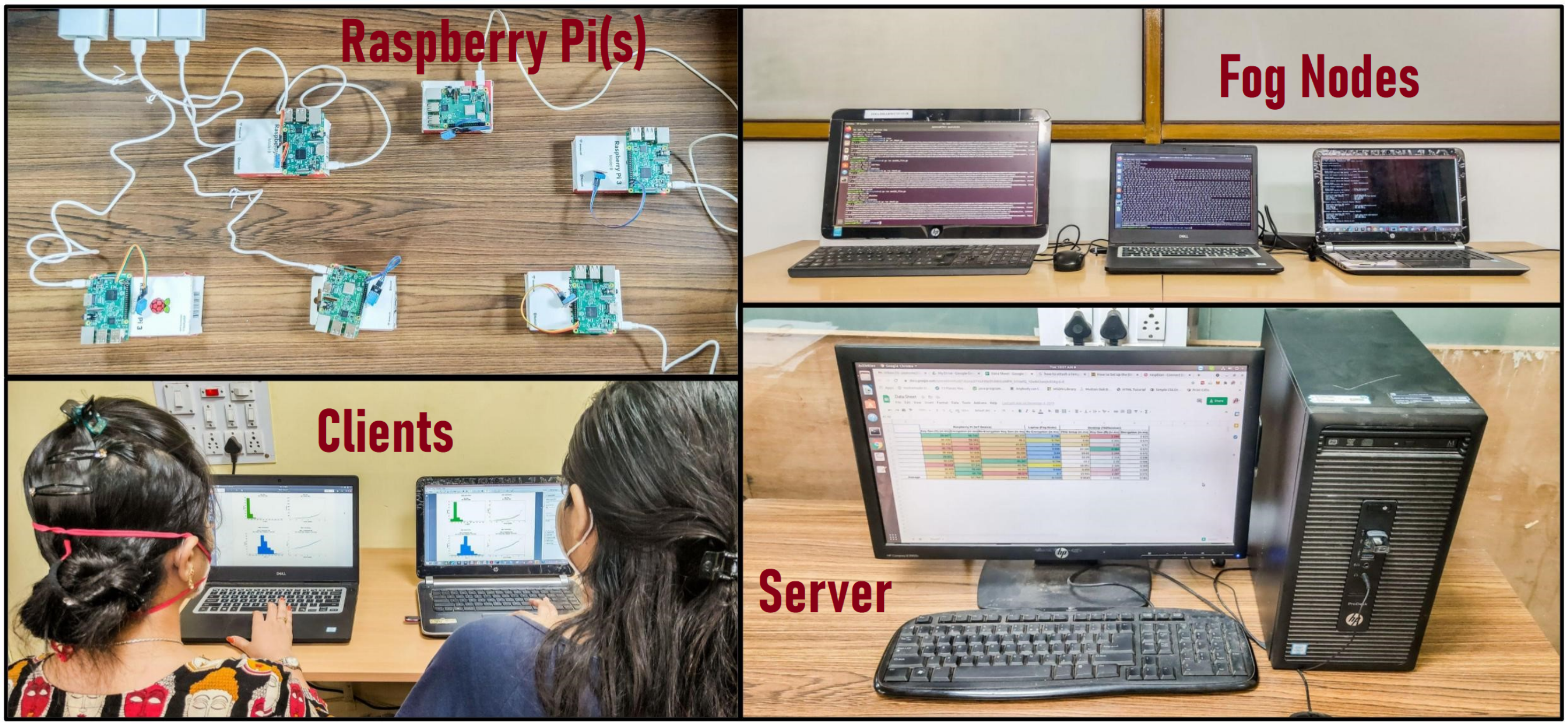}
\setlength{\belowcaptionskip}{-10pt}
\caption{\small \sl Testbed Setup of \textit{FairShare} \label{fig:Image9}}
\end{center} 
\end{figure}

\noindent \textbf{Testbed Setup:} Experimentation is conducted through testbed where Raspberry Pi(s) (RPI-3B) act as the Industrial IoT Device (D). We have deployed six Raspberry Pi(s) which are equipped with DHT11 temperature, humidity sensors. Multiple laptops serve as Fog Nodes and a desktop acts as the server in our setup. Finally, a couple of clients operating at a remote terminal retrieve data from the system. Fig. \ref{fig:Image9} shows our testbed set-up. The specifications of devices used in the testbed are shown in Table \ref{tab:Table6}. The same codes used for simulation have been deployed. Through our experimentation, we examine the performance of \textit{FairShare} based on the various evaluation metrics described in Section \ref{em}.

\begin{table}[!htb]
\centering
\caption{\small \sl Testbed Setup Specifications}
\label{tab:Table6}
\scalebox{0.62}{
\begin{tabular}{c|c|c|c|c}
\hline
\rowcolor[HTML]{9B9B9B} 
\textbf{Specifications} & \textbf{\begin{tabular}[c]{@{}c@{}}Industrial IoT\\ Device/s (D)\end{tabular}} & \textbf{Fog Node/s (F)} & \textbf{Server} & \textbf{Client/s (C)} \\ \hline
Memory & 1 GBB & 3.8 GiB & 7.7 GiB & 3.5 GiB \\ \hline
Processor & \begin{tabular}[c]{@{}c@{}}Cortex-A53, armv7l\\ @1200MHz * 4\end{tabular} & \begin{tabular}[c]{@{}c@{}}Intel® Core™ i5-7200U\\ CPU @ 2.50GHz * 4\end{tabular} & \begin{tabular}[c]{@{}c@{}}Intel® Core™ i7-6700\\ CPU @ 3.40GHz * 8\end{tabular} & \begin{tabular}[c]{@{}c@{}}Intel® Core™ i5-4460T\\ CPU @ 1.90GHz * 4\end{tabular} \\ \hline
OS & 32-bit Raspbian & \begin{tabular}[c]{@{}c@{}}64-bit Ubuntu\\ 18.04.3\end{tabular} & \begin{tabular}[c]{@{}c@{}}64-bit Ubuntu\\ 18.04.3\end{tabular} & \begin{tabular}[c]{@{}c@{}}64-bit Ubuntu\\ 18.04.3\end{tabular} \\ \hline
Disk & 16 GB & 455.1 GB & 983.4 GB & 100.3 GB \\ \hline
\end{tabular}%
}
\end{table}

\begin{table*}[!t]
\begin{minipage}[b]{0.30\linewidth}
\centering
\caption{\small \sl Experimental Results}
\label{tab:Table7}
\scalebox{0.57}{
\begin{tabular}{c|c|c}
\hline
\rowcolor[HTML]{9B9B9B} 
\textbf{Devices} & \textbf{Tasks} & \textbf{Computation Time (ms)} \\ \hline
 & Key Generation & 3307.797 \\ \cline{2-3} 
 & Signature Generation & 57.307 \\ \cline{2-3} 
 & Symmetric Encryption ($c_m$) & 0.174 \\ \cline{2-3} 
\multirow{-4}{*}{IIoT Device (D)} & Asymmetric Encryption ($c_k$) & 2.933 \\ \hline
 & Key Generation & 378.649 \\ \cline{2-3} 
 & Asymmetric Decryption of $c_k$ & 1.577 \\ \cline{2-3} 
 & Symmetric Decryption of $c_m$ & 0.002 \\ \cline{2-3} 
\multirow{-4}{*}{Fog Node (F)} & Signature Verification & 0.073 \\ \hline
\end{tabular}
}
\end{minipage}
\hspace{0.3cm}
\begin{minipage}[b]{0.62\linewidth}
\centering
\caption{\small \sl Overheads related to different Blockchain calls}
\label{tab:Table8}
\scalebox{0.61}{
\begin{tabular}{l|c|c|c|c|c|c|c|c}
\hline
\multicolumn{1}{c|}{\multirow{2}{*}{\textbf{\begin{tabular}[c]{@{}c@{}}Type\\ of Task\end{tabular}}}} & \multirow{2}{*}{\textbf{Tasks}} & \multicolumn{3}{c|}{\textbf{Execution Time (sec)}} & \multirow{2}{*}{\textbf{\begin{tabular}[c]{@{}c@{}}Gas\\ Usage\end{tabular}}} & \multirow{2}{*}{\textbf{\begin{tabular}[c]{@{}c@{}}Deployment\\ Cost (ETH)\end{tabular}}} & \multirow{2}{*}{\textbf{\begin{tabular}[c]{@{}c@{}}Deployment\\ Cost (USD)\end{tabular}}} & \multirow{2}{*}{\textbf{\begin{tabular}[c]{@{}c@{}}Communication\\ Overhead (bytes)\end{tabular}}} \\ \cline{3-5}
\multicolumn{1}{c|}{} &  & \textit{\textbf{One Fog Node}} & \textit{\textbf{Two Fog Nodes}} & \textit{\textbf{Three Fog Nodes}} &  &  &  &  \\ \hline
\multirow{4}{*}{\textbf{Deploy}} & Store $nonce$ in $\mathbf{S}$ & 4.748 & 7.646 & 10.079 & 1078 & 0.000037730 & 0.11 & 12 \\ \cline{2-9} 
 & Store ${Meta}_1$ in $\mathbf{S}$ & 3.414-11.273* & 3.789-11.598* & 4.656-13.392* & 1725 & 0.000060375 & 0.17 & 68 \\ \cline{2-9} 
 & Store Access Policy in $\mathbf{S}$ & 3.413 & 5.428 & 7.812 & 956 & 0.000033460 & 0.095 & 6 \\ \cline{2-9} 
 & Store ${Meta}_2$ in $\mathbf{S}$ & 5.015 & 6.695 & 8.012 & 1698 & 0.000059430 & 0.17 & 36 \\ \hline
\multirow{2}{*}{\textbf{Retrieve}} & $Compare\_AccessPolicy()$ & 0.000890082 & Same & Same & - & - & - & 1 \\ \cline{2-9} 
 & $Verify\_KeyHash()$ & 0.0007871038 & Same & Same & - & - & - & 1 \\ \hline
\end{tabular}
}
\end{minipage}
\end{table*}

\noindent \textbf{Results and Discussion} In our testbed experimentation, the interactions between IIoT device (D) and Fog node (F) which were previously omitted during simulation are also included. We conduct four additional set of experiments and the average results obtained after ten independent runs are recorded.

In the first set of experiment, we plot (Fig. \ref{fig:Image10}) time taken to deploy metadata related to a file in contract $\mathbf{S}$ with varying file size in the presence of multiple fog nodes. From the figure, we observe that blockchain execution time grows linearly till 500KB file size, followed by an exponential growth. As explained earlier, x-axis values in our experimentation are not linear, therefore execution time growth looks exponential for larger files. Despite blockchain execution time being a few seconds, it wouldn't affect the performance of the IIoT system in the long run because for a single file $f$ to be shared with different clients storing metadata about $f$ in $\mathbf{S}$ is a one-time task. Additionally, we observe that execution time grows linearly with increasing number of fog nodes. This is because number of blockchain calls increases proportionately with increasing number of fog nodes operating in unison.

In the second set of experiment, Table \ref{tab:Table7} shows the tasks performed during the interactions between D and F for the purpose of data generation. We observe that the computation times of each of the tasks lie within a feasible range and is therefore suitable for IIoT scenarios.

\begin{figure}[!b]
    \begin{minipage}[t]{.47\linewidth}
        \centering
        \includegraphics[scale=0.20]{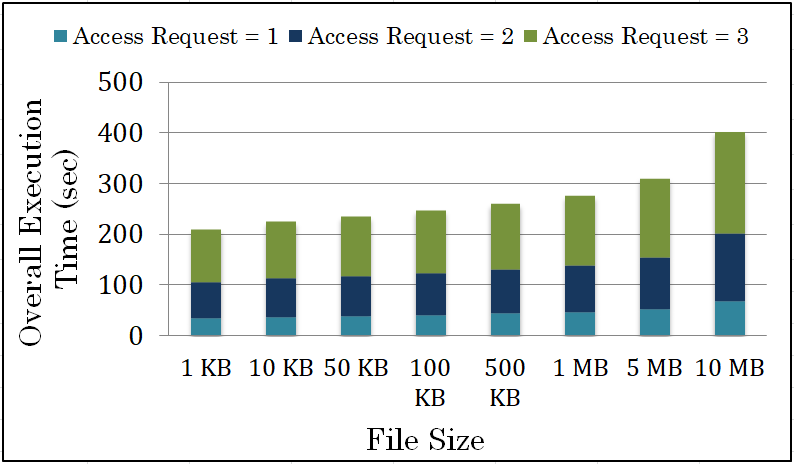}
        \subcaption{Overall Execution Time for different $|F|$} \label{fig:Image6a}
    \end{minipage}
   \hspace{0.3cm}
    \begin{minipage}[t]{.46\linewidth}
        \centering
        \includegraphics[scale=0.215]{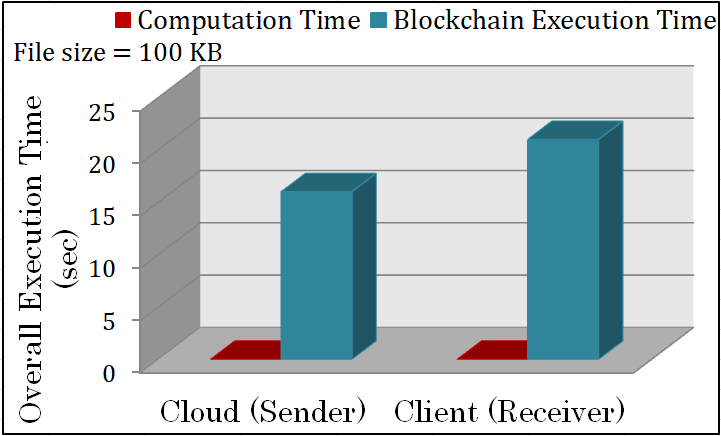}
        \subcaption{Overall Execution Time for each Party}\label{fig:Image6b}
    \end{minipage}
    \label{fig:Image6}
    \setlength{\belowcaptionskip}{-10pt}
    \caption{\small \sl Fair Exchange of File between Cloud and Client}
\end{figure}

In the third set of experiment, we evaluate the various parameters related to fair exchange of the requested file ($f$) between cloud (Sender) and client (Receiver). Firstly, we plot (Fig. \ref{fig:Image6a}) the overall execution time for fair exchange of $f$ with varying size. In this plot, we also take into consideration the number of file requests made by multiple clients. We observe from the figure that the variation in overall execution time is almost linear throughout the plot, except for a slight spike for 10MB file size. The linear trend occurs because the time consuming tasks in this case are certain blockchain calls related to the fair exchange of file. These blockchain call times are independent of the file size. Thus, the variations in overall execution time is caused due to the computation time of the parties involved, which varies with file size. Additionally, we observe that overall execution time grows linearly with increase in number of file access requests made. This is because, our testbed setup uses a single CPU as the server which processes requests sequentially. However, real life cloud servers are equipped with vast computer clusters which leverage parallel processing and would thus reduce execution time significantly. In the subsequent Fig. \ref{fig:Image6b}, we plot overall execution time for each party involved in communication for a fixed file size of 100KB. From the figure, we observe the computation time for the individual parties is very negligible $\approx 0.005\%$ (represented as flat red boxes in the plot) in comparison to blockchain execution time. This is because fair exchange of file involves a lot of blockchain related operations which are computationally intensive compared to other tasks due to the inherent overhead of blockchain.

Finally, in the last experiment, we show (Table \ref{tab:Table8}) overheads related to various smart contract based blockchain calls. The per unit gas price as well as its equivalent value in USD is calculated at the same exchange rate as mentioned earlier. We observe from the table that the deployment related calls are more expensive compared to ones used for fetching data from the blockchain.

We summarily observe from the entire testbed experimentation that the execution time grows linearly in the presence of multiple fog and client nodes. This is because the number of blockchain calls is proportional to either the number of file access requests made or the number of fog nodes operating in unison. We also observe that \textit{FairShare} performs considerably well up to file sizes of 1MB. However, for larger file sizes (i.e. beyond 1MB), the overheads are slightly higher which may be a limitation but it comes at the cost of achieving accountability and fairness while still guaranteeing fair exchange of file. Currently, \textit{FairShare} allows accessing only non-real-time data, thus it is not well suited for time-critical IIoT applications which is another limitation that we plan to address in future. 

\section{Conclusion \label{Section7}}

The \textit{FairShare} preserves accountability and fairness while ensuring security and privacy of the exchanged data within minimal overheads by utilising proxy re-encryption. It also ensures data access by authorised entities only against some fee by integrating smart contracts into the blockchain. Through a detailed security analysis, we prove that our scheme efficiently handles collusion attacks under certain adversarial assumptions. Moreover, results obtained through prototype implementation of \textit{FairShare} have shown that our scheme is feasible, enhances data sharing process without the requirement of centralized trust and scales considerably well even for larger file sizes.  

\noindent As future extension, the scheme may be extended by including resource optimization and load balancing in fog nodes.  Also, guaranteeing data privacy by examining more security threats is another open area that requires attention. 

\bibliographystyle{IEEEtran}
{\footnotesize
\bibliography{references}}

\begin{thebibliography}{10}
\providecommand{\url}[1]{#1}
\csname url@samestyle\endcsname
\providecommand{\newblock}{\relax}
\providecommand{\bibinfo}[2]{#2}
\providecommand{\BIBentrySTDinterwordspacing}{\spaceskip=0pt\relax}
\providecommand{\BIBentryALTinterwordstretchfactor}{4}
\providecommand{\BIBentryALTinterwordspacing}{\spaceskip=\fontdimen2\font plus
\BIBentryALTinterwordstretchfactor\fontdimen3\font minus
  \fontdimen4\font\relax}
\providecommand{\BIBforeignlanguage}[2]{{%
\expandafter\ifx\csname l@#1\endcsname\relax
\typeout{** WARNING: IEEEtran.bst: No hyphenation pattern has been}%
\typeout{** loaded for the language `#1'. Using the pattern for}%
\typeout{** the default language instead.}%
\else
\language=\csname l@#1\endcsname
\fi
#2}}
\providecommand{\BIBdecl}{\relax}
\BIBdecl

\bibitem{i3}
S.~Z. {Tajalli}, M.~{Mardaneh}, E.~{Taherian-Fard}, A.~{Izadian},
  A.~{Kavousi-Fard}, M.~{Dabbaghjamanesh}, and T.~{Niknam}, ``{DoS-Resilient
  Distributed Optimal Scheduling in a Fog Supporting IIoT-Based Smart
  Microgrid},'' \emph{IEEE Transactions on Industry Applications}, vol.~56,
  no.~3, pp. 2968--2977, 2020.

\bibitem{i2}
J.~{Sengupta}, S.~{Ruj}, and S.~D. {Bit}, ``{A Secure Fog-Based Architecture
  for Industrial Internet of Things and Industry 4.0},'' \emph{IEEE
  Transactions on Industrial Informatics}, vol.~17, no.~4, pp. 2316--2324,
  2021.

\bibitem{i4}
J.~{Tsai}, I.~{Chuang}, J.~{Liu}, Y.~{Kuo}, and W.~{Liao}, ``{QoS-Aware Fog
  Service Orchestration for Industrial Internet of Things},'' \emph{IEEE
  Transactions on Services Computing}, 2020.

\bibitem{i6}
D.~A. {Chekired}, L.~{Khoukhi}, and H.~T. {Mouftah}, ``{Industrial IoT Data
  Scheduling Based on Hierarchical Fog Computing: A Key for Enabling Smart
  Factory},'' \emph{IEEE Transactions on Industrial Informatics}, vol.~14,
  no.~10, pp. 4590--4602, 2018.

\bibitem{XWZ_ASIACCS'12}
L.~Xu, X.~Wu, and X.~Zhang, ``{CL-PRE: A Certificateless Proxy Re-encryption
  Scheme for Secure Data Sharing with Public Cloud},'' in \emph{Proceedings of
  the 7th ACM Symposium on Information, Computer and Communications Security},
  ser. ASIACCS '12, 2012.

\bibitem{r12}
C.~H. {Liu}, Q.~{Lin}, and S.~{Wen}, ``{Blockchain-Enabled Data Collection and
  Sharing for Industrial IoT With Deep Reinforcement Learning},'' \emph{IEEE
  Transactions on Industrial Informatics}, vol.~15, no.~6, pp. 3516--3526,
  2019.

\bibitem{r21}
Q.~{Wen}, Y.~{Gao}, Z.~{Chen}, and D.~{Wu}, ``{A Blockchain-based Data Sharing
  Scheme in The Supply Chain by IIoT},'' in \emph{2019 IEEE International
  Conference on Industrial Cyber Physical Systems (ICPS)}.\hskip 1em plus 0.5em
  minus 0.4em\relax IEEE, 2019, pp. 695--700.

\bibitem{r13}
X.~{Zheng} and Z.~{Cai}, ``{Privacy-Preserved Data Sharing Towards Multiple
  Parties in Industrial IoTs},'' \emph{IEEE Journal on Selected Areas in
  Communications}, vol.~38, no.~5, pp. 968--979, 2020.

\bibitem{r18}
A.~{Dorri}, A.~{Hill}, S.~{Kanhere}, R.~{Jurdak}, F.~{Luo}, and Z.~Y. {Dong},
  ``{Peer-to-Peer EnergyTrade: A Distributed Private Energy Trading
  Platform},'' in \emph{2019 IEEE International Conference on Blockchain and
  Cryptocurrency (ICBC)}.\hskip 1em plus 0.5em minus 0.4em\relax IEEE, 2019,
  pp. 61--64.

\bibitem{r22}
V.~{Hassija}, V.~{Chamola}, S.~{Garg}, D.~N.~G. {Krishna}, G.~{Kaddoum}, and
  D.~N.~K. {Jayakody}, ``{A Blockchain-Based Framework for Lightweight Data
  Sharing and Energy Trading in V2G Network},'' \emph{IEEE Transactions on
  Vehicular Technology}, vol.~69, no.~6, pp. 5799--5812, 2020.

\bibitem{r14}
Y.~{Lu}, X.~{Huang}, Y.~{Dai}, S.~{Maharjan}, and Y.~{Zhang}, ``{Blockchain and
  Federated Learning for Privacy-Preserved Data Sharing in Industrial IoT},''
  \emph{IEEE Transactions on Industrial Informatics}, vol.~16, no.~6, pp.
  4177--4186, 2020.

\bibitem{r5}
S.~{Cui}, M.~R. {Asghar}, and G.~{Russello}, ``{Towards Blockchain-Based
  Scalable and Trustworthy File Sharing},'' in \emph{2018 27th International
  Conference on Computer Communication and Networks (ICCCN)}.\hskip 1em plus
  0.5em minus 0.4em\relax IEEE, 2018, pp. 1--2.

\bibitem{r6}
D.~{Li}, R.~{Du}, Y.~{Fu}, and M.~H. {Au}, ``{Meta-Key: A Secure Data-Sharing
  Protocol Under Blockchain-Based Decentralized Storage Architecture},''
  \emph{IEEE Networking Letters}, vol.~1, no.~1, pp. 30--33, 2019.

\bibitem{r7}
A.~{Manzoor}, M.~{Liyanage}, A.~{Braeke}, S.~S. {Kanhere}, and M.~{Ylianttila},
  ``{Blockchain based Proxy Re-Encryption Scheme for Secure IoT Data
  Sharing},'' in \emph{2019 IEEE International Conference on Blockchain and
  Cryptocurrency (ICBC)}.\hskip 1em plus 0.5em minus 0.4em\relax IEEE, 2019,
  pp. 99--103.

\bibitem{r9}
B.~{Sharma}, R.~{Halder}, and J.~{Singh}, ``{Blockchain-based Interoperable
  Healthcare using Zero-Knowledge Proofs and Proxy Re-Encryption},'' in
  \emph{2020 International Conference on COMmunication Systems NETworkS
  (COMSNETS)}.\hskip 1em plus 0.5em minus 0.4em\relax IEEE, 2020, pp. 1--6.

\bibitem{r10}
S.~Badsha, I.~Vakilinia, and S.~Sengupta, ``{BloCyNfo-Share: Blockchain based
  Cybersecurity Information Sharing with Fine Grained Access Control},'' in
  \emph{10th Annual Computing and Communication Workshop and Conference, {CCWC}
  2020, Las Vegas, NV, USA, January 6-8, 2020}.\hskip 1em plus 0.5em minus
  0.4em\relax {IEEE}, 2020, pp. 317--323.

\bibitem{r23}
B.~Chen, D.~He, N.~Kumar, H.~Wang, and K.~R. Choo, ``{A Blockchain-Based Proxy
  Re-Encryption With Equality Test for Vehicular Communication Systems},''
  \emph{IEEE Transactions on Network Science and Engineering}, vol.~8, no.~3,
  pp. 2048--2059, 2021.

\bibitem{r25}
Y.~Gao, Y.~Chen, X.~Hu, H.~Lin, Y.~Liu, and L.~Nie, ``{Blockchain Based IIoT
  Data Sharing Framework for SDN-Enabled Pervasive Edge Computing},''
  \emph{IEEE Transactions on Industrial Informatics}, vol.~17, no.~7, pp.
  5041--5049, 2021.

\bibitem{r26}
A.~Manzoor, A.~Braeken, S.~S. Kanhere, M.~Ylianttila, and M.~Liyanage, ``{Proxy
  re-encryption enabled secure and anonymous IoT data sharing platform based on
  blockchain},'' \emph{Journal of Network and Computer Applications}, vol. 176,
  p. 102917, 2021.

\bibitem{PG99}
H.~Pagnia and F.~C. Gärtner, ``{On the Impossibility of Fair Exchange without
  a Trusted Third Party},'' Tech. Rep., 1999.

\bibitem{r15}
P.~Gupta, S.~Kanhere, and R.~Jurdak, ``{A Decentralized IoT Data
  Marketplace},'' \emph{CoRR}, vol. abs/1906.01799, 2019.

\bibitem{r16}
R.~Xu, G.~S. Ramachandran, Y.~Chen, and B.~Krishnamachari, ``{BlendSM-DDM:
  BLockchain-ENabled Secure Microservices for Decentralized Data
  Marketplaces},'' in \emph{2019 IEEE International Smart Cities Conference
  (ISC2)}.\hskip 1em plus 0.5em minus 0.4em\relax IEEE, 2019, pp. 14--17.

\bibitem{r17}
G.~S. {Ramachandran}, R.~{Radhakrishnan}, and B.~{Krishnamachari}, ``{Towards a
  Decentralized Data Marketplace for Smart Cities},'' in \emph{2018 IEEE
  International Smart Cities Conference (ISC2)}.\hskip 1em plus 0.5em minus
  0.4em\relax IEEE, 2018, pp. 1--8.

\bibitem{r24}
H.~{Xu}, Q.~{He}, X.~{Li}, B.~{Jiang}, and K.~{Qin}, ``{BDSS-FA: A
  Blockchain-Based Data Security Sharing Platform With Fine-Grained Access
  Control},'' \emph{IEEE Access}, vol.~8, pp. 87\,552--87\,561, 2020.

\bibitem{r20}
S.~{Matsumoto} and R.~M. {Reischuk}, ``{IKP: Turning a PKI Around with
  Decentralized Automated Incentives},'' in \emph{2017 IEEE Symposium on
  Security and Privacy (SP)}.\hskip 1em plus 0.5em minus 0.4em\relax IEEE,
  2017, pp. 410--426.

\bibitem{r19}
E.~{Wagner}, A.~{Völker}, F.~{Fuhrmann}, R.~{Matzutt}, and K.~{Wehrle},
  ``{Dispute Resolution for Smart Contract-based Two-Party Protocols},'' in
  \emph{2019 IEEE International Conference on Blockchain and Cryptocurrency
  (ICBC)}.\hskip 1em plus 0.5em minus 0.4em\relax IEEE, 2019, pp. 422--430.

\bibitem{c3}
S.~Dziembowski, L.~Eckey, and S.~Faust, ``{FairSwap: How To Fairly Exchange
  Digital Goods},'' in \emph{Proceedings of the 2018 ACM SIGSAC Conference on
  Computer and Communications Security}, ser. CCS '18.\hskip 1em plus 0.5em
  minus 0.4em\relax ACM, 2018, pp. 967--984.

\bibitem{AK_ICBC'19}
A.~Asgaonkar and B.~Krishnamachari, ``{Solving the Buyer and Seller’s
  Dilemma: A Dual-Deposit Escrow Smart Contract for Provably Cheat-Proof
  Delivery and Payment for a Digital Good without a Trusted Mediator},'' in
  \emph{2019 IEEE International Conference on Blockchain and Cryptocurrency
  (ICBC)}, 2019, pp. 262--267.

\bibitem{SMHK_JPDC'21}
P.~Singh, M.~Masud, M.~S. Hossain, and A.~Kaur, ``{Cross-domain secure data
  sharing using blockchain for industrial IoT},'' \emph{Journal of Parallel and
  Distributed Computing}, vol. 156, pp. 176--184, 2021.

\bibitem{r27}
J.~{Xie}, H.~{Tang}, T.~{Huang}, F.~R. {Yu}, R.~{Xie}, J.~{Liu}, and Y.~{Liu},
  ``{A Survey of Blockchain Technology Applied to Smart Cities: Research Issues
  and Challenges},'' \emph{IEEE Communications Surveys Tutorials}, vol.~21,
  no.~3, pp. 2794--2830, 2019.

\bibitem{c10}
J.~Sengupta, S.~Ruj, and S.~Das~Bit, ``{An Efficient and Secure Directed
  Diffusion in Industrial Wireless Sensor Networks},'' in \emph{Proceedings of
  the 1st International Workshop on Future Industrial Communication Networks
  (FICN), MobiCom}.\hskip 1em plus 0.5em minus 0.4em\relax ACM, 2018, p.
  41–46.

\bibitem{c4}
D.~Boneh and M.~K. Franklin, ``{Identity-Based Encryption from the Weil
  Pairing},'' in \emph{Advances in Cryptology - {CRYPTO}}, vol. 2139.\hskip 1em
  plus 0.5em minus 0.4em\relax Springer, 2001, pp. 213--229.

\bibitem{c2}
B.~Libert and D.~Vergnaud, ``{Unidirectional Chosen-Ciphertext Secure Proxy
  Re-encryption},'' in \emph{Public Key Cryptography -- PKC 2008}.\hskip 1em
  plus 0.5em minus 0.4em\relax Springer, 2008, pp. 360--379.

\bibitem{c6}
J.~Sengupta, S.~Ruj, and S.~D. Bit, ``{A Comprehensive Survey on Attacks,
  Security Issues and Blockchain Solutions for IoT and IIoT},'' \emph{Journal
  of Network and Computer Applications}, vol. 149, p. 102481, 2020.

\bibitem{c8}
G.~Wood, ``{Ethereum: A secure decentralised generalised transaction ledger},''
  \emph{Ethereum project yellow paper}, vol. 151, pp. 1--32, 2014.

\bibitem{c17}
K.~Peng, M.~Li, H.~Huang, C.~Wang, S.~Wan, and K.-K.~R. Choo, ``{Security
  Challenges and Opportunities for Smart Contracts in Internet of Things: A
  Survey},'' \emph{IEEE Internet of Things Journal}, vol.~8, no.~15, pp.
  12\,004--12\,020, 2021.

\bibitem{c5}
P.~Banerjee, N.~Nikam, and S.~Ruj, ``{Blockchain Enabled Privacy Preserving
  Data Audit},'' \emph{CoRR}, vol. abs/1904.12362, 2019.

\bibitem{WSB_ANTS'19}
S.~Walling, J.~Sengupta, and S.~Das~Bit, ``{A Low-cost Real-time IoT based Air
  Pollution Monitoring using LoRa},'' in \emph{IEEE International Conference on
  Advanced Networks and Telecommunications Systems (ANTS)}, 2019, pp. 1--6.

\bibitem{SRB_CCGrid'22}
J.~Sengupta, S.~Ruj, and S.~D. Bit, ``{SPRITE: A Scalable Privacy-Preserving
  and Verifiable Collaborative Learning for Industrial IoT},'' in \emph{22nd
  IEEE International Symposium on Cluster, Cloud and Internet Computing
  (CCGrid)}, 2022, pp. 249--258.

\bibitem{c16}
M.~A. Ferrag and L.~Shu, ``{The Performance Evaluation of Blockchain-based
  Security and Privacy Systems for the Internet of Things: A Tutorial},''
  \emph{IEEE Internet of Things Journal}, vol.~8, no.~24, pp. 17\,236--17\,260,
  2021.

\bibitem{HET_BC'19}
J.~Heiss, J.~Eberhardt, and S.~Tai, ``{From Oracles to Trustworthy Data
  On-Chaining Systems},'' in \emph{2019 IEEE International Conference on
  Blockchain (Blockchain)}, 2019, pp. 496--503.

\bibitem{PV_ICC'20}
M.~Pincheira and M.~Vecchio, ``{Towards Trusted Data on Decentralized IoT
  Applications: Integrating Blockchain in Constrained Devices},'' in \emph{2020
  IEEE International Conference on Communications Workshops (ICC Workshops)},
  2020, pp. 1--6.

\bibitem{c14}
C.~Dong, Y.~Wang, A.~Aldweesh, P.~McCorry, and A.~van Moorsel, ``{Betrayal,
  Distrust, and Rationality: Smart Counter-Collusion Contracts for Verifiable
  Cloud Computing},'' in \emph{Proceedings of the 2017 ACM SIGSAC Conference on
  Computer and Communications Security}.\hskip 1em plus 0.5em minus 0.4em\relax
  ACM, 2017, p. 211–227.

\bibitem{NC_22}
L.~J\"{a}ger, D.~Lorych, and M.~Eckel, ``{A Resilient Network Node for the
  Industrial Internet of Things},'' in \emph{Proceedings of the 17th
  International Conference on Availability, Reliability and Security}, 2022.

\bibitem{r1}
Y.~Liu, Y.~Ren, C.~Ge, J.~Xia, and Q.~Wang, ``{A CCA-secure Multi-Conditional
  Proxy Broadcast Re-encryption Scheme for Cloud Storage System},''
  \emph{Journal of Information Security and Applications}, vol.~47, pp. 125 --
  131, 2019.

\bibitem{r2}
C.~Ge, Z.~Liu, J.~Xia, and L.~Fang, ``{Revocable Identity-Based Broadcast Proxy
  Re-Encryption for Data Sharing in Clouds},'' \emph{IEEE Transactions on
  Dependable and Secure Computing}, vol.~18, no.~3, pp. 1214--1226, 2021.

\bibitem{r4}
P.~Shabisha, A.~Braeken, and K.~Steenhaut, ``{Symmetric Key-Based Secure
  Storage and Retrieval of IoT Data on a Semi-trusted Cloud Server},''
  \emph{Wireless Personal Communications}, vol. 113, no.~1, pp. 537--553, 2020.

\bibitem{l5}
B.~Lynn, ``{PBC Library : The Pairing-Based Cryptography Library},''
  \url{https://crypto.stanford.edu/pbc/}, 2013.

\bibitem{c12}
M.~Saqlain, M.~Piao, Y.~Shim, and J.~Lee, ``{Framework of an IoT-based
  Industrial Data Management for Smart Manufacturing},'' \emph{Journal of
  Sensor and Actuator Networks}, vol.~8, no.~25, 2019.

\bibitem{c13}
S.~Chen, X.~Zhu, H.~Zhang, C.~Zhao, G.~Yang, and K.~Wang, ``{Efficient Privacy
  Preserving Data Collection and Computation Offloading for Fog-Assisted
  IoT},'' \emph{IEEE Transactions on Sustainable Computing}, vol.~5, no.~4, pp.
  526--540, 2020.

\end{thebibliography}

\appendices

\section{Symmetric Key Encryption \label{AppA}}

A symmetric key encryption technique (e.g. AES) consists of the following three functions:

\begin{itemize}[leftmargin=*]
    \item $Sym.KeyGen() \rightarrow (K_{id})$ : Generates a symmetric key to be shared between two parties.
    \item $Sym.Enc(K_{id} , m) \rightarrow (c)$ : Takes as input plaintext message $m$ and symmetric key $K_{id}$ to generate ciphertext $c$.
    \item $Sym.Dec(K_{id} , c) \rightarrow (m)$ : Performs the steps exactly in the reverse order as performed by $Sym.Enc()$ to retrieve the plaintext message using the same key $K_{id}$.
\end{itemize}

\section{Asymmetric Key Encryption \label{AppB}}

A public-key encryption scheme is a triple of probabilistic polynomial-time algorithms as follows:

\begin{itemize}[leftmargin=*]
    \item $Asym.KeyGen (params) \rightarrow (sk^1_{id},pk^1_{id})$ : Generates a private/public key-pair for each of the parties.
    \item $Asym.Enc (pk^1_{id} , m) \rightarrow (c)$ : Takes as input a message $m$ and public key $pk^1_{id}$ to generate ciphertext $c$.
    \item $Dec2 (sk^1_{id} , c) \rightarrow (m)$ : Derives plaintext message $m$ using private key $sk^1_{id}$ and ciphertext $c$ as inputs.
\end{itemize}

\section{Detailed Working Principle of \textit{FairShare} \label{AppC}}

\noindent \textbf{Working Principle:} Our proposed scheme is broken down into three major parts. The details of each of these parts are outlined as follows:

\noindent \textbf{Part 1 : Initialization}

\noindent In this part, each of the parties involved in the communication converge on the security parameters $<\lambda, \mathbb{G}_1, \mathbb{G}_T, g, q>$ and derive their necessary keys as below:
\begin{enumerate}[label=(\alph*),leftmargin=*]
    \item The devices in the perception layer (IIoT Device D)
    generate their own symmetric key $K_D$ (\textit{Appendix \ref{AppA}}) and a key-pair ($sk^1_D , pk^1_D$) using $Asym.KeyGen (params)$ (\textit{Appendix \ref{AppB}}).
    \item Next, Fog node (F) derives it's own symmetric key $K_F$ (\textit{Appendix \ref{AppA}}) and a key-pair ($sk_F , pk_F$) for performing re-encryption using $PRE.KeyGen(\lambda, par)$ (Section II-B). It also computes another key-pair ($sk^1_F , pk^1_F$) using $Asym.KeyGen (params)$ (\textit{Appendix \ref{AppB}}).
    \item Lastly, Client (C) derives it's key-pair ($sk_C , pk_C$) using $PRE.KeyGen(\lambda, par)$ (Section II-B).
\end{enumerate}

\noindent \textbf{Part 2 : Data Generation to Storage}

\noindent To effectively store an encrypted data in Cloud, so that it can later be retrieved by the concerned parties, the following steps need to be executed :

\noindent\textit{\textbf{Step 1:}} Whenever device $D$ has some sensed data $m_i$, it performs the following:

\begin{enumerate}[label=(\alph*),leftmargin=*]
    \item It first encrypts the data packet using its symmetric key [here $h=H(m_i)$ and  $\sigma=\mathcal{S}(sk^1_D,h)$]
    $$c_m = Sym.Enc_{K_D}(m_i||h|| \sigma||DId)$$
    \item Next, D encrypts the symmetric key ($K_D$) with the Fog node's public key i.e. $c_k = Asym.Enc_{pk^1_F}(K_D)$. Lastly, D forwards the 2-tuple $<c_m,c_k>$ to the destined F.
\end{enumerate}

\noindent\textit{\textbf{Step 2:}} After receiving the 2-tuple from D, F does the following:

\begin{enumerate}[leftmargin=*, label=(\alph*)]
    \item It first performs $Asym.Dec_{sk^1_F}(Asym.Enc_{{pk^1}_F}(K_D))$ to derive the symmetric key $K_D$.
    \item F then uses this symmetric key to decrypt $Sym.Dec_{K_D}(Sym.Enc_{K_D}(m_i||h|| \sigma||DId))$ in order to derive the packet ($m||h|| \sigma||DId$). 
    \item F computes $h'=H(m_i)$ and compares it with the received $h$. On successful verification, F is assured that the integrity of $m_i$ is preserved. F then computes $\mathcal{V}(\sigma,pk^1_D,m_i)$ to verify the authenticity of the data source. On successful verification, F is further assured that $m_i$ has indeed been sent by device D and the following step gets executed. If either of the above verification fails, the entire process is \textit{aborted}.
\end{enumerate}

\noindent\textit{\textbf{Step 3:}} On successful verification, this step begins where F provides a feedback to D for each data $m_i$ it receives and then stores the summaries of these data to a file $f$. This file is then encrypted using a symmetric key $K_F$ to generate $f'=Sym.Enc_{K_F}(f)$ before uploading it to Cl. To reduce additional overheads and introduce flexibility, \textit{FairShare} implements proxy re-encyption on key $K_F$ instead of applying it on file $f$. So, for effectively sharing the symmetric key ($K_F$), F performs the following:

\begin{enumerate}[label=(\alph*),leftmargin=*]
    \item It randomly chooses $r \in \mathbb{Z}^*_q$ and computes $c_1$ and $c_2$ as:
$$c_1 = ({pk_F})^r \ ;\  c_2 = \hat{e}(g,g)^r.K_F$$
\noindent where $g$ is the generator of the bilinear map groups. This step produces a ciphertext computed over key $K_F$, that can later be re-encrypted again.
    \item F maintains a table which stores final encrypted key $c'=(c_1,c_2)$ respective to each file. It further computes the meta data related to $f'$ and $c'$ as $Meta_1=(DId||h_1||h_2)$ and stores it as a transaction in blockchain B [here, $h_1=H(f')$ and $h_2=H(c')$].
    \item Finally, the access policies ($\rho$) with respect to $f/f'$ are stored in a Controller Contract ($\mathbf{S}$).
\end{enumerate}

\noindent\textit{\textbf{Step 4:}} The cloud on receiving the encrypted file $f'$ computes $h'_1 = H(f')$ and compares it with $h_1$ stored as $Meta_1$ in B. On successful verification, Cl is assured that integrity of received $f'$ is preserved and the scheme is ready for execution of \textbf{Part 3}. On failure, Cl concludes that $f'$ has been sent incorrectly or tampered midway, and it thus \textit{aborts} the entire process.

\noindent \textbf{Part 3: Data Request to Retrieval}

\noindent At any point of time, when Client (C) sends a request $Req$ to access $f$ along with an incentive $\epsilon_1$ to Cl, the following steps are performed: 

\noindent\textit{\textbf{Step 1:}} Initially, Cl calls $\mathbf{S}$ to verify whether C has access rights to $f$. On successful verification, the following steps are executed, else the process is \textit{aborted}.  

\noindent\textit{\textbf{Step 2:}} Cl forwards the $Req$ along with an incentive $\epsilon_2$ to F. Simultaneously, it also creates Judge Contract ($\mathbf{J}$) to securely share $f'$ with C. Now, for sharing $f'$ with the client, C locks $P_1$ coins as the price of $f'$ and Cl locks $P_2$ coins as a safety deposit respectively in $\mathbf{J}$. In our scheme, the predicate $\phi$ is the hash of the encrypted file $f'$ to be shared with C. Finally, Cl initiates the fair exchange of $f'$ with C.

\noindent Here, the steps 3 and 5 are executed in parallel by the concerned parties and are followed by their succeeding steps.

\noindent\textit{\textbf{Step 3:}} Meanwhile, after receiving $<Req,\epsilon_2>$ from Cl, F generates re-encryption key ($rk_{F \rightarrow C} = {pk_C}^{1/{sk_F}}$) and does the following:

\begin{enumerate}[label=(\alph*),leftmargin=*]
    \item  F randomly chooses $t \in \mathbb{Z}^*_q$ and re-encrypts $c'$ under $rk_{F \rightarrow C}$ as follows:

$$c'_1 = {rk^{1/t}_{F \rightarrow C}}\ ;\ c''_1 = c^t_1$$
Therefore, the re-encrypted ciphertext is $c''=(c'_1,c''_1,c_2)$. This re-encrypted version of key $K_F$ can later be decrypted by the Client to retrieve $K_F$ which can then be used for further decryption $f'$.
    \item F also stores the meta data related to $c''$, $Meta_2 = (DId||h_3)$ as a transaction in B [here, $h_3=H(c'')$].
    \item Finally, for secure exchange of $c''$, C locks $P_3$ coins in $\mathbf{S}$ as the price of $c''$. Similarly, F locks $P_4$ coins as the safety deposit in $\mathbf{S}$ and initiates the exchange of $c''$.
\end{enumerate}

\noindent\textit{\textbf{Step 4:}} On receiving $c''$, C first computes $h'_3 = H(c'')$ and verifies it with $h_3$ stored as $Meta_2$ in B. On successful verification, $P_3 +P_4$ coins are transferred to F as the price for sharing $c''$ with C. If the hash verification fails, it means that F is \textbf{malicious} and has sent an incorrect $c''$. In such a case, F's safety deposit is seized, $P_3 +P_4$ coins are transferred back to C and the process is \textit{aborted}.

\noindent Otherwise, $c''$ is decrypted using $sk_C$ to retrieve the shared key $K_F$:
$$K_F=c_2/\hat{e}(c'_1,c''_1)^{1/sk_C}$$

\noindent This shared key $K_F$ that was originally used to encrypt file $f$ to $f'$ can now be used to decrypt $f'$ to retrieve the original file $f$.

\noindent\textit{\textbf{Step 5:}} One of the following three test case scenarios can arise after C receives $f'$ :

\begin{enumerate}[label=(\alph*),leftmargin=*]
    \item If C claims the received file to be valid, then, $P_1 + P_2$ coins are transferred to Cl as a price for sharing $f'$ and the scheme proceeds to \textbf{Step 6}. 
    \item Otherwise, if C generates a complain against Cl with a valid proof of misbehavior, a verification of the proof is carried out by $\mathbf{J}$. $\mathbf{J}$ then detects Cl to be \textbf{malicious} for sending an incorrect $f'$ and hence seizes its safety deposit. In such a case, $P_1 + P_2$ coins are transferred to C and the process aborted.
    \item Lastly, if either of the two scenarios doesn't occur, then after a specific timestamp $ts$, Cl calls $\mathbf{J}$ to trigger the release of $P_1 + P_2$ coins in its favour.
    
\end{enumerate}

\noindent\textit{\textbf{Step 6:}} Finally, C decrypts $f'$ using $K_F$ to retrieve $f = Sym.Dec_{K_F}(f')$ and the scheme successfully terminates.

\vskip -2\baselineskip plus -1fil

\begin{IEEEbiography}[{\includegraphics[width=1in,height=1.25in,clip,keepaspectratio]{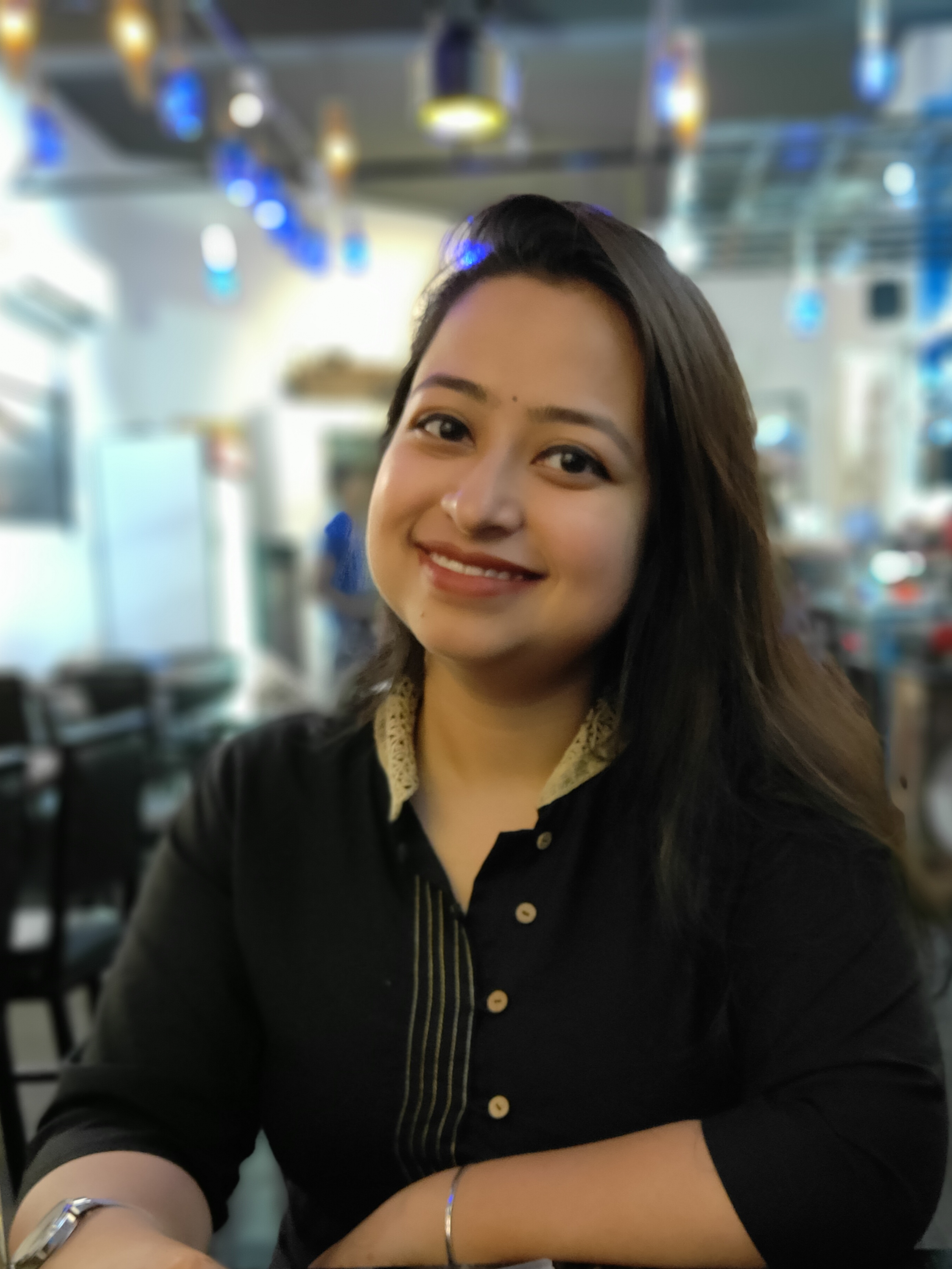}}]{Jayasree Sengupta} is currently working as a Postdoctoral Researcher at CISPA Helmholtz Center for Information Security, Germany. In 2022, she has completed her PhD in Computer Science from Indian Institute of Engineering Science and Technology, Shibpur, India. She received her MTech degree in Distributed and Mobile Computing from Jadavpur University, India in 2017. She has published research articles in reputed peer reviewed journals and International conference proceedings. Her research interests include Applied Cryptography, Blockchains, Fog computing, IoT/IIoT and Data Privacy.
\end{IEEEbiography}

\vskip -2\baselineskip plus -1fil

\begin{IEEEbiography}[{\includegraphics[width=1in,height=1.25in,clip,keepaspectratio]{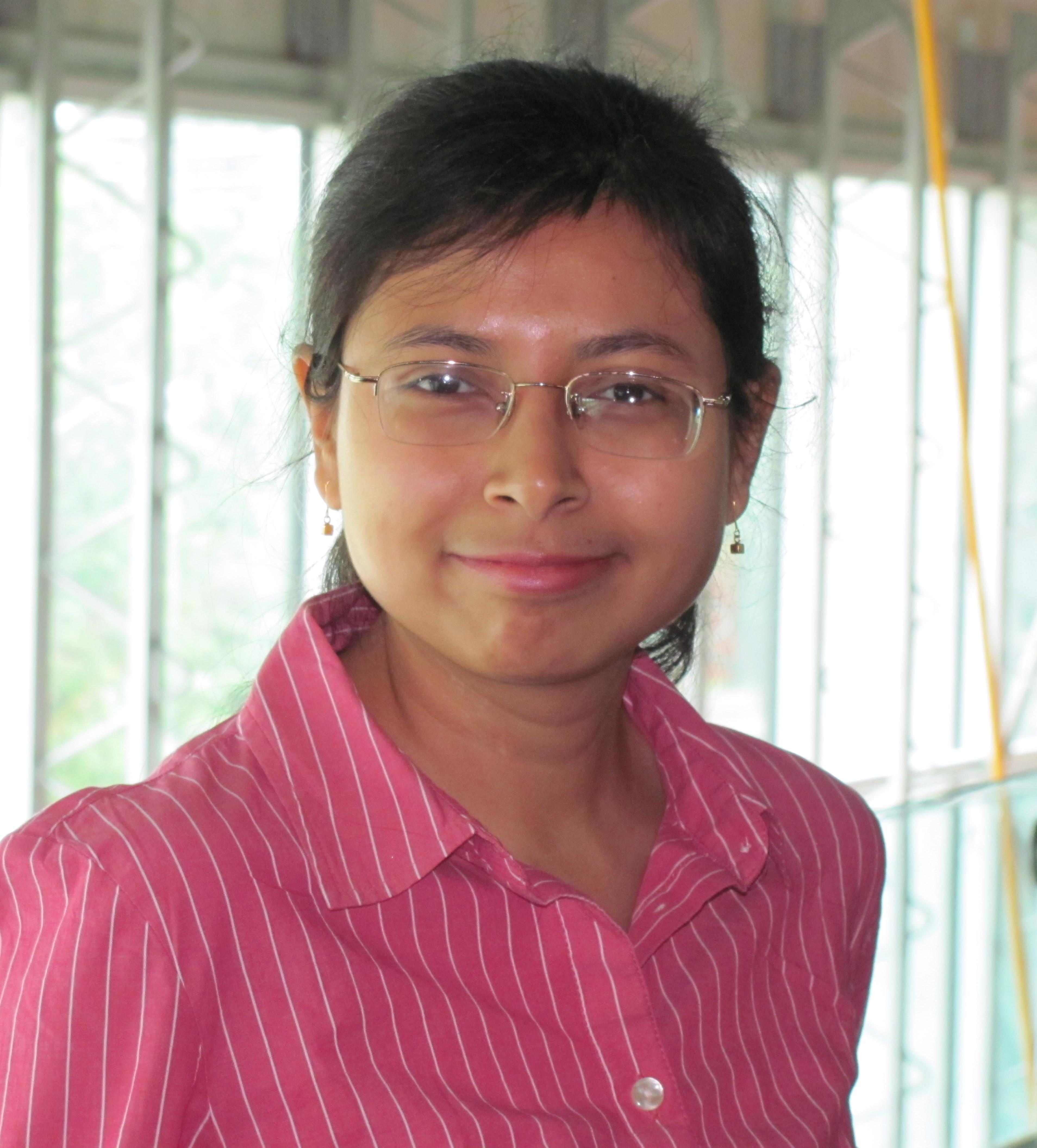}}]{Sushmita Ruj} is a Senior Lecturer at University of New South Wales, Sydney, Australia. Prior to that, she had served as a Senior Research Scientist at CSIRO Data61, Australia and as an Associate Professor at ISI, Kolkata. She was an Erasmus Mundus Post-Doctoral Fellow at Lund University, Sweden and Post-Doctoral Fellow at University of Ottawa, Canada. Her research interests are in Blockchains, Applied Cryptography, and Data Privacy. She is a recipient of Samsung GRO award, NetApp Faculty Fellowship, Cisco Academic Grant and IBM OCSP grant. She is a Senior Member of ACM and IEEE.
\end{IEEEbiography}


\vskip -2\baselineskip plus -1fil

\begin{IEEEbiography}[{\includegraphics[width=1in,height=1.25in,clip,keepaspectratio]{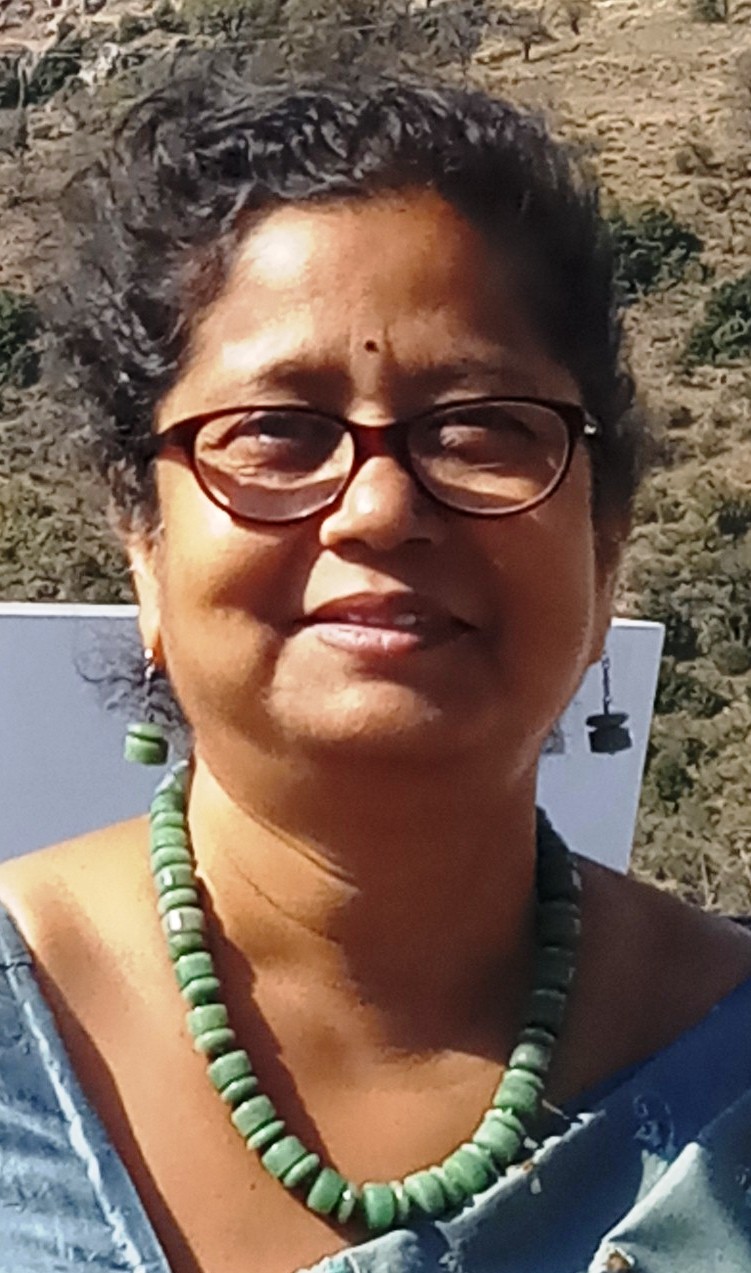}}]{Sipra Das Bit}
 is a Professor of the Department of Computer Science and Technology, Indian Institute of Engineering Science and Technology, Shibpur, India. She also served as a visiting professor in the Department of Information and Communication Technology, Asian Institute of Technology, Bangkok in 2017. A recipient of the Career Award for Young Teachers from the All India Council of Technical Education (AICTE), she has more than 30 years of teaching and research experience. Professor Das Bit has published many research papers in reputed journals and refereed International conference proceedings. She also has three books to her credit. Her current research interests include Internet of Things, wireless sensor network, delay tolerant network, mobile computing and network security. She is a Senior Member of IEEE.
\end{IEEEbiography}

\end{document}